\documentclass[preprint2,numberedappendix,tighten,twocolappendix]{aastex6}
\shorttitle{Hydrogen Epoch of Reionization Array (HERA)}
\shortauthors{DeBoer, et al.}

\usepackage{amsmath}
\usepackage{graphicx}
\usepackage[figuresright]{rotating}
\usepackage{natbib}
\usepackage{ctable}
\usepackage{tabularx}
\usepackage{enumitem}
\usepackage{url}
\setlist[itemize]{noitemsep, topsep=0pt}
\setlist[enumerate]{noitemsep, topsep=0pt}
\newcommand{\Mycitet}[1]{\citet{#1}}
\newcommand{\Mycitep}[1]{\citep{#1}}

\newcommand{\kvec}{{\bf k}}

\newcommand{\kpr}{{k_\perp}}
\newcommand{\kvpr}{{\kvec_\perp}}

\newcommand{\integral}{\int\limits}

\newcommand{\Caption}[4]{\vspace{#1}\renewcommand{\baselinestretch}{#2}\caption{#4}\vspace{#3}}
\def\kperp{k_{\bot}}
\def\eppsilon{{$\varepsilon$ppsilon}}
\def\kpar{k_{\|}}

\begin{document}
\title{Hydrogen Epoch of Reionization Array (HERA)}

\author{David R. DeBoer\altaffilmark{1}, Aaron R. Parsons\altaffilmark{1}, James E. Aguirre\altaffilmark{10}, Paul  Alexander\altaffilmark{5}, Zaki S. Ali\altaffilmark{1}, Adam P. Beardsley\altaffilmark{2}, Gianni  Bernardi\altaffilmark{11, 13}, Judd D. Bowman\altaffilmark{2}, Richard F. Bradley\altaffilmark{8}, Chris L. Carilli\altaffilmark{9}, Carina  Cheng\altaffilmark{1}, Eloy  de~Lera~Acedo\altaffilmark{5}, Joshua S. Dillon\altaffilmark{1}, Aaron  Ewall-Wice\altaffilmark{7}, Gcobisa  Fadana\altaffilmark{13}, Nicolas  Fagnoni\altaffilmark{5}, Randall  Fritz\altaffilmark{13}, Steve R. Furlanetto\altaffilmark{4}, Brian  Glendenning\altaffilmark{9}, Bradley  Greig\altaffilmark{12}, Jasper  Grobbelaar\altaffilmark{13}, Bryna J. Hazelton\altaffilmark{14, 15}, Jacqueline N. Hewitt\altaffilmark{7}, Jack  Hickish\altaffilmark{1}, Daniel C. Jacobs\altaffilmark{2}, Austin  Julius\altaffilmark{13}, MacCalvin  Kariseb\altaffilmark{13}, Saul A. Kohn\altaffilmark{10}, Telalo  Lekalake\altaffilmark{13}, Adrian  Liu\altaffilmark{1, 16}, Anita  Loots\altaffilmark{13}, David  MacMahon\altaffilmark{1}, Lourence  Malan\altaffilmark{13}, Cresshim  Malgas\altaffilmark{13}, Matthys  Maree\altaffilmark{13}, Zachary  Martinot\altaffilmark{10}, Nathan  Mathison\altaffilmark{13}, Eunice  Matsetela\altaffilmark{13}, Andrei  Mesinger\altaffilmark{12}, Miguel F. Morales\altaffilmark{14}, Abraham R. Neben\altaffilmark{7}, Nipanjana  Patra\altaffilmark{1}, Samantha  Pieterse\altaffilmark{13}, Jonathan C. Pober\altaffilmark{3}, Nima  Razavi-Ghods\altaffilmark{5}, Jon  Ringuette\altaffilmark{14}, James  Robnett\altaffilmark{9}, Kathryn  Rosie\altaffilmark{13}, Raddwine  Sell\altaffilmark{13}, Craig  Smith\altaffilmark{13}, Angelo  Syce\altaffilmark{13}, Max  Tegmark\altaffilmark{7}, Nithyanandan  Thyagarajan\altaffilmark{2}, Peter K.~G. Williams\altaffilmark{6}, Haoxuan  Zheng\altaffilmark{7}}
\altaffiltext{1}{Department of Astronomy, University of California, Berkeley, CA}
\altaffiltext{2}{School of Earth and Space Exploration, Arizona State University, Tempe, AZ}
\altaffiltext{3}{Physics Department, Brown University, Providence, RI}
\altaffiltext{4}{Department of Physics and Astronomy, University of California, Los Angeles, CA}
\altaffiltext{5}{Cavendish Astrophysics, University of Cambridge, Cambridge, UK}
\altaffiltext{6}{Harvard-Smithsonian Center for Astrophysics, Cambridge, MA}
\altaffiltext{7}{Department of Physics, Massachusetts Institute of Technology, Cambridge, MA}
\altaffiltext{8}{National Radio Astronomy Observatory, Charlottesville, VA}
\altaffiltext{9}{National Radio Astronomy Observatory, Socorro, NM}
\altaffiltext{10}{Department of Physics and Astronomy, University of Pennsylvania, Philadelphia, PA}
\altaffiltext{11}{Department of Physics and Electronics, Rhodes University, PO Box 94, Grahamstown, 6140, South Africa}
\altaffiltext{12}{Scuola Normale Superiore, Pisa, Italy}
\altaffiltext{13}{SKA-SA, Cape Town, South Africa}
\altaffiltext{14}{Department of Physics, University of Washington, Seattle, WA}
\altaffiltext{15}{eScience Institute, University of Washington, Seattle, WA}
\altaffiltext{16}{Hubble Fellow}

\email{ddeboer@berkeley.edu}

\begin{abstract}
The Hydrogen Epoch of Reionization Array (HERA) is a staged experiment to measure 21 cm emission from the primordial intergalactic medium (IGM) throughout cosmic reionization ($z=6-12$), and to explore earlier epochs of our Cosmic Dawn ($z\sim30$).  During these epochs, early stars and black holes heated and ionized the IGM, introducing fluctuations in 21 cm emission.  HERA is designed to characterize the evolution of the 21 cm power spectrum to constrain the timing and morphology of reionization, the properties of the first galaxies, the evolution of large-scale structure, and the early sources of heating.   The full HERA instrument will be a 350-element interferometer in South Africa consisting of 14-m parabolic dishes observing from 50 to 250 MHz.  Currently, 19 dishes have been deployed on site  and the next 18 are under construction.  HERA has been designated as an SKA Precursor instrument. 
In this paper, we summarize HERA's scientific context and provide forecasts for its key science results.  After reviewing the current state of the art in foreground mitigation, we use the delay-spectrum technique to motivate high-level performance requirements for the HERA instrument.  Next, we present the HERA instrument design, along with the subsystem specifications that ensure that HERA meets its performance requirements.  Finally, we summarize the schedule and status of the project.  We conclude by suggesting that, given the realities of foreground contamination, current-generation 21 cm instruments are approaching their sensitivity limits. HERA is designed to bring both the sensitivity and the precision to deliver its primary science on the basis of proven foreground filtering techniques, while developing new subtraction techniques to unlock new capabilities.  The result will be a major step toward realizing the widely recognized scientific potential of 21 cm cosmology.
\end{abstract}


\section{Introduction}

The Hydrogen Epoch of Reionization Array (HERA; http://reionization.org) is a staged experiment to use the redshifted 21\,cm line
of neutral hydrogen to characterize our Cosmic Dawn, from
the formation of the first stars and black holes $\sim0.1~$Gyr after the Big
Bang ($z\sim30$) through the full reionization of the intergalactic medium
(IGM) $\sim1~$Gyr later ($z\sim6$).  By directly observing the large scale
structure of the primordial IGM as it is heated and reionized, HERA complements
probes at other wavelengths, adding transformative capabilities for
understanding the astrophysics and fundamental cosmology of our early universe.
Taking advantage of a new understanding of bright foreground systematics, HERA's
purpose-built radio
interferometer is optimized to deliver high signal-to-noise measurements of
redshifted 21\,cm emission to detect and characterize the Epoch of Reionization (EOR).

The power that observations of highly redshifted hydrogen emission have for answering key science about our early universe
has motivated a resurgence of interest in low-frequency arrays --- most with the primary
objective of measuring the EOR.  These include the Precision Array Probing the Epoch of Reionization (PAPER; \citealt{parsons_et_al2010}), the Murchison Widefield Array (MWA; \citealt{tingay_et_al2013}), the LOw Frequency ARray (LOFAR; \citealt{2013A&A...556A...2V}), and the Long Wavelength Array (LWA; \citealt{2009IEEEP..97.1421E}), as well as systems for existing dish arrays like the Very Large Array (VLA; \citealt{2013AAS...22132804K}) and the Giant Metrewave Radio Telescope (GMRT; \citealt{paciga_et_al2011}).
These experiments all struggle with the
inherent challenge of simultaneously meeting stringent sensitivity requirements
while suppressing foregrounds $\sim 5-6$ orders of magnitude
brighter than the 21\,cm signal \citep[][]{bernardi_et_al2009,pober_etal2013b,dillon_et_al2014}.
HERA improves on its predecessors by bringing significantly more sensitivity to bear
on the angular and spectral scales where recent work (discussed in Section \ref{sec:delayapproach}) has indicated that
the power spectrum of the EoR may dominate over foregrounds.

\begin{figure*}
	\centering
	\includegraphics[height=1.6in]{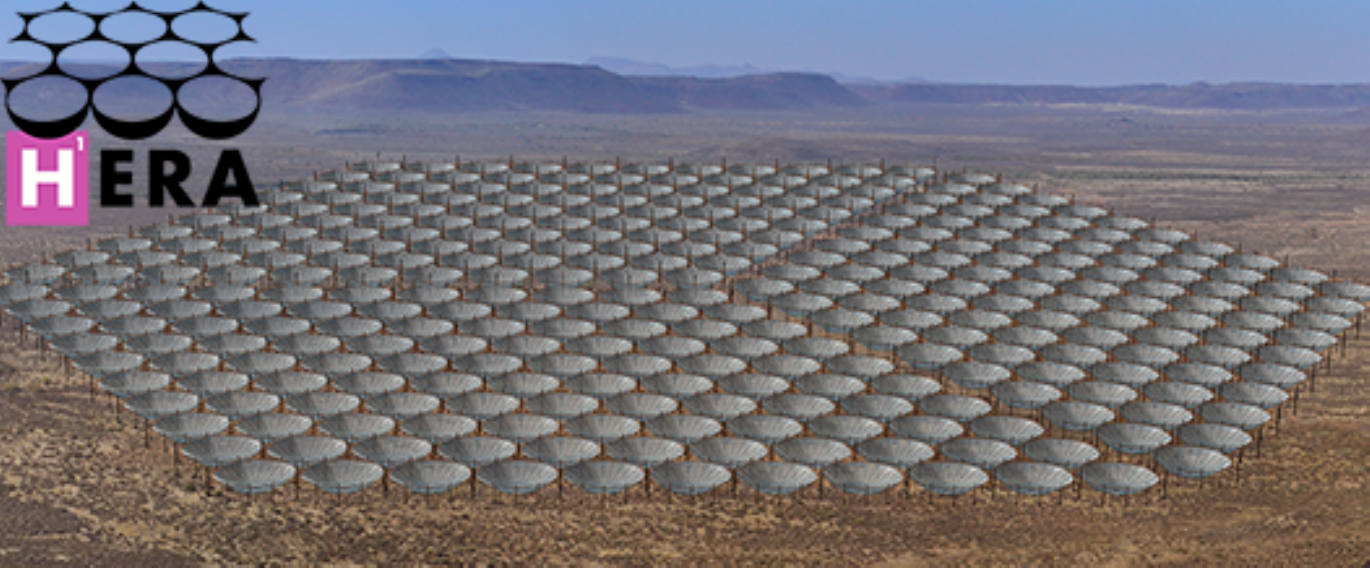}
	\includegraphics[height=1.6in]{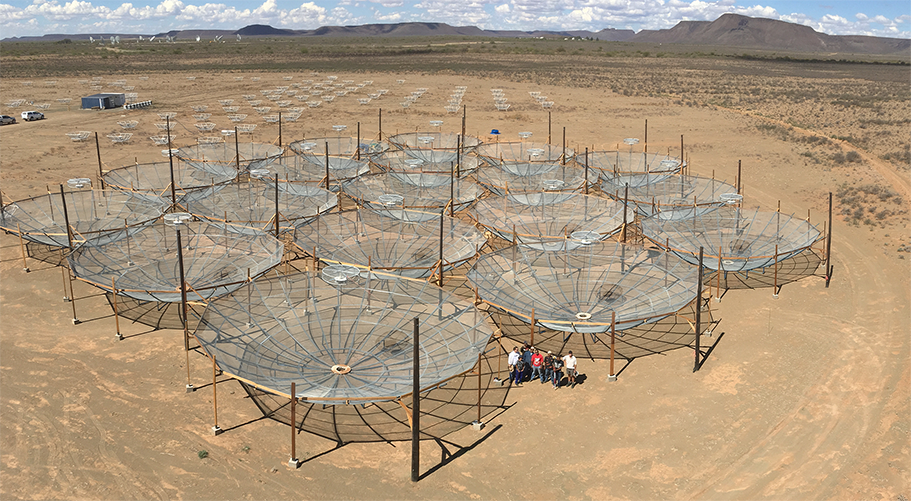}
	\caption{Rendering of the 320-element core (left) of the full HERA-350 array and picture of 19 HERA 14-m, zenith-pointing dishes (with PAPER elements in the background) currently deployed in South Africa (right).} 
	\label{fig:HERApictures}
\end{figure*}

The HERA experiment will comprise 
350 14-m parabolic dishes (320 in a dense core $+$ 30 outriggers) in the South African Karoo Radio
Astronomy Reserve (see Fig. \ref{fig:HERApictures}, top, for a rendering).
HERA's antenna element and its compact configuration are
optimized for robust power spectrum detection,
delivering the requisite collecting area inexpensively, while carefully controlling foreground systematics.
The design and construction of this instrument is supported by the HERA collaboration 
consisting of the following Partner institutions:  
Arizona State University (Tempe, AZ USA), 
Brown University (Providence, RI USA), 
University of California Berkeley (Berkeley, CA USA), 
University of California Los Angeles (Los Angeles, CA USA), 
University of Cambridge (Cambridge, UK), 
Massachusetts Institute of Technology (Cambridge, MA USA), 
National Radio Astronomy Observatory (Charlottesville, VA USA), 
University of Pennsylvania (Philadelphia, PA USA), 
Scuola Normale Superiore di Pisa (Pisa, Italy),
SKA-South Africa (Cape Town, South Africa), and
University of Washington (Seattle, WA USA).  
Additional Collaborators are at 
Harvard University (Cambridge, MA USA),  
University of KwaZulu Natal (Durban, South Africa), 
University of Western Cape (Cape Town, South Africa), 
Imperial College London  (London, UK) and 
California State Polytechnic University (Pomona, CA USA).
The South African National Research Foundation Square Kilometre Array South Africa  (SKA-SA) group is a key partner in HERA's construction and science.

HERA's first stage of development has been funded under the US National Science Foundation's Mid-Scale Innovations Program,
which has supported the construction of a 19-element array for testing instrumental performance
and manufacturability on location.  The first 19-element array is complete and, with additional funding from Cambridge University, 
construction is now underway on the next 18 elements.
The total 37 elements, to be completed in 2016, will provide a factor of about 5 more 
sensitivity than PAPER-128 and provide a significant chance at detecting the EoR power spectrum signal.  The next proposed phase will build out to 128 elements and use the existing 128 dual-polarization analog and digital signal paths that have been in use with PAPER.
HERA-128 should provide a robust detection of the EOR signal and allow some characterization.  Finally, extending to 2020, HERA will build out to 350 elements to further EOR science as a function 
of redshift and spatial scale, potentially producing the first images of the EOR.  
As a low-frequency science array on an SKA site, HERA has been designated as an SKA Precursor instrument.

This paper is structured as follows.  Section \ref{sec:science} gives a high-level overview of the science motivating HERA's construction, including primary and secondary science objectives. 
Section \ref{sec:eormeas} presents the techniques used to make the measurement.  Section \ref{sec:requirements} presents the high-level requirements and Section \ref{sec:design} presents the system description.   Section \ref{sec:status} presents a brief status and outline the deployement timetable and Section \ref{sec:conclusion} concludes by summarizing and providing additional context.

\section{Scientific Background}
\label{sec:science}

\begin{figure}
	\centering
	\includegraphics[width=0.48\textwidth]{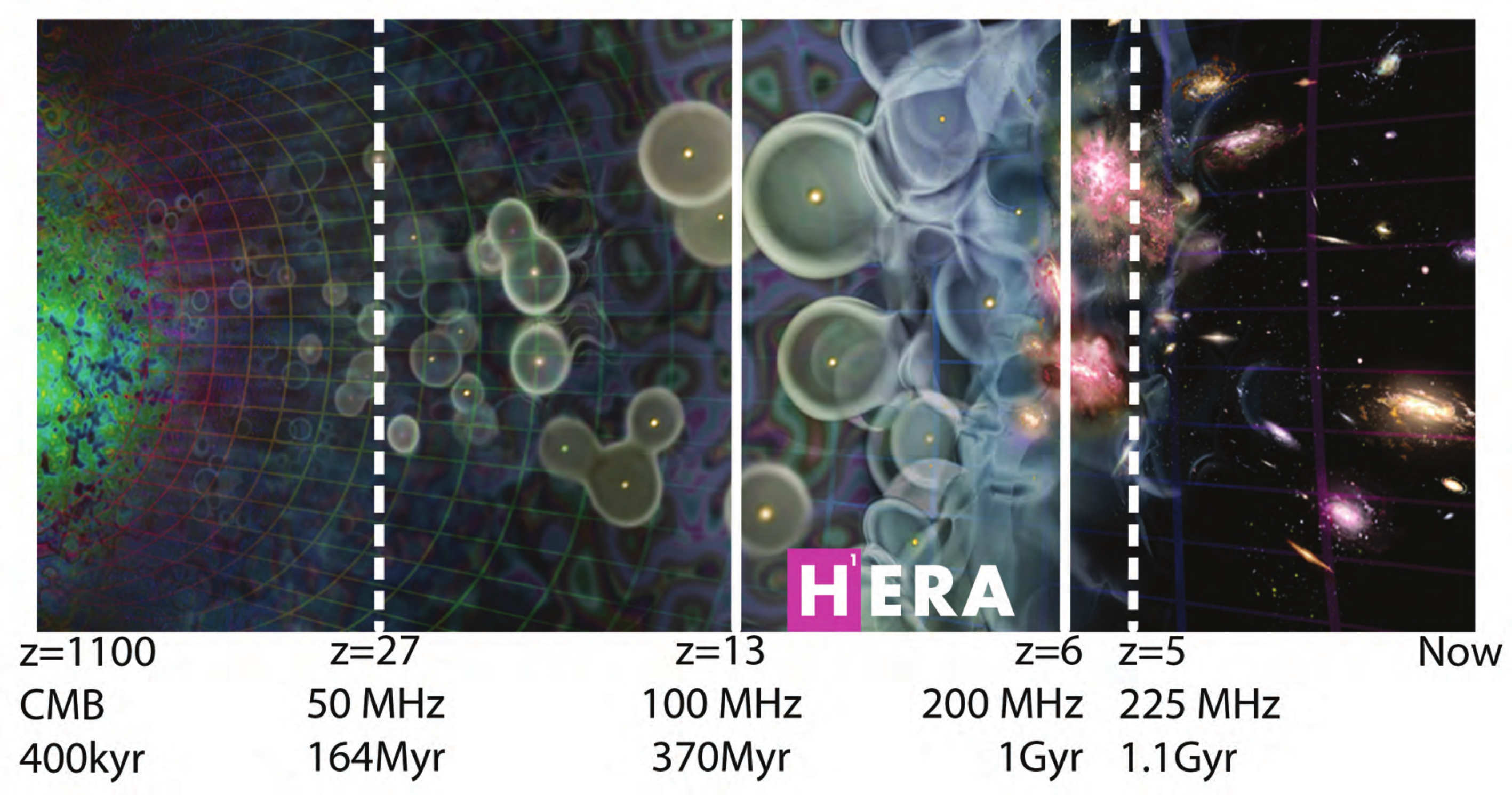}
	\caption{Rendering of cosmic evolution from just after the Big Bang to today (background image credit Loeb/{\em Scientific American}).  The labels show the redshift and the frequency of the red-shifted hydrogen line (rest frequency 1420 MHz) at different ages of the Universe.  The solid white lines bracket the HERA EOR band and the dashed ones bracket the extended frequency goal.  
CMB observations observe the afterglow of the Big Bang (far left) and Baryonic Acoustic Oscillation (BAO) surveys proposed 
target $z\approx0.8$-2.5.  Limited surveys span back to about $z\approx7$.}
	\label{fig:cosmos}
\end{figure}

The {\it Cosmic Dawn} of our universe is one of the last unexplored
frontiers in cosmic history.  This history is summarized in Figure~\ref{fig:cosmos}, starting with the Big Bang on the left. The hot, young universe expands and cools slowly in the background while gravitational instability around concentrations of dark matter causes primordial density fluctuations to grow.
The Cosmic Dawn represents a specific epoch in this growth, where the first stars and galaxies formed and illuminated the Universe en route to forming the astronomical structures we see today. Ultimately, this early population of gravitationally condensed material produced sufficiently energetic flux to reionize the intergalactic medium (IGM) from its previous neutral state in a period called the Epoch of Reionization.  This period is shown in Figure~\ref{fig:cosmos} as the rapid transition from separated large reionized ``bubbles'' to the merged reionized state and to structures that begin to resemble the denizens of our current universe. The structure of the 
IGM thus contains a panoply of information about the underlying astrophysical and cosmological phenomena governing cosmic evolution.

The evolution of the cosmic structure depends on the local and average cosmic density, the relative velocities of
baryons and dark matter, and the sizes and clustering of the first galaxies to form.
But it also depends on the constituents of those first galaxies---so-called Population
III stars (stars formed very early with little to no elements heavier than helium), later generation stars, stellar remnants, X-ray binaries,
and early supermassive black holes.  Bulk properties like
ultraviolet and X-ray luminosities and spectra also affect the thermal and
ionization states of the IGM.  The wealth of unexplored physics during the Cosmic Dawn,
culminating in the Epoch of Reionization, 
led the most recent US National Academies astronomy decadal survey entitled {\em New Worlds, New Horizons} to highlight it as one of the top three ``priority science objectives'' for
the decade \citep{NWNH}.

Exploring the interplay of galaxies and large-scale structure during the EOR
requires complementary observational approaches. Measurements of the Cosmic Microwave Background 
(CMB; the photons permeating the universe after becoming
transparent to its own radiation by the recombination of the protons and electrons about 400,000 years after the Big Bang)  by COBE, WMAP and Planck provide initial conditions for structure formation.  Thomson
scattering of CMB photons by the ionized particles constrains the integrated column of 
ionized gas and kinetic Sunyaev-Zel'dovich measurements constrain the duration of the ``patchy'' phase of cosmic structures.
But even with these measurements, the detailed evolution of the IGM 
is only loosely constrained \citep{haiman_holder2003,mortonson_hu2008,zahn_etal2012,mesinger_et_al2012}.
Lyman-$\alpha$ absorption features in quasar and $\gamma$-ray burst spectra give 
ionization constraints at the tail end of reionization 
($z<7$, \citealt{fan_et_al2006, mcgreer_et_al2015}), but these features 
saturate at low neutral fractions
$x_{\rm HI} \ga 10^{-4}$, where $x_{\rm HI}$ is the fraction of hydrogen in its neutral state. 

\begin{figure}
\centering
    \includegraphics[width=0.48\textwidth,clip]{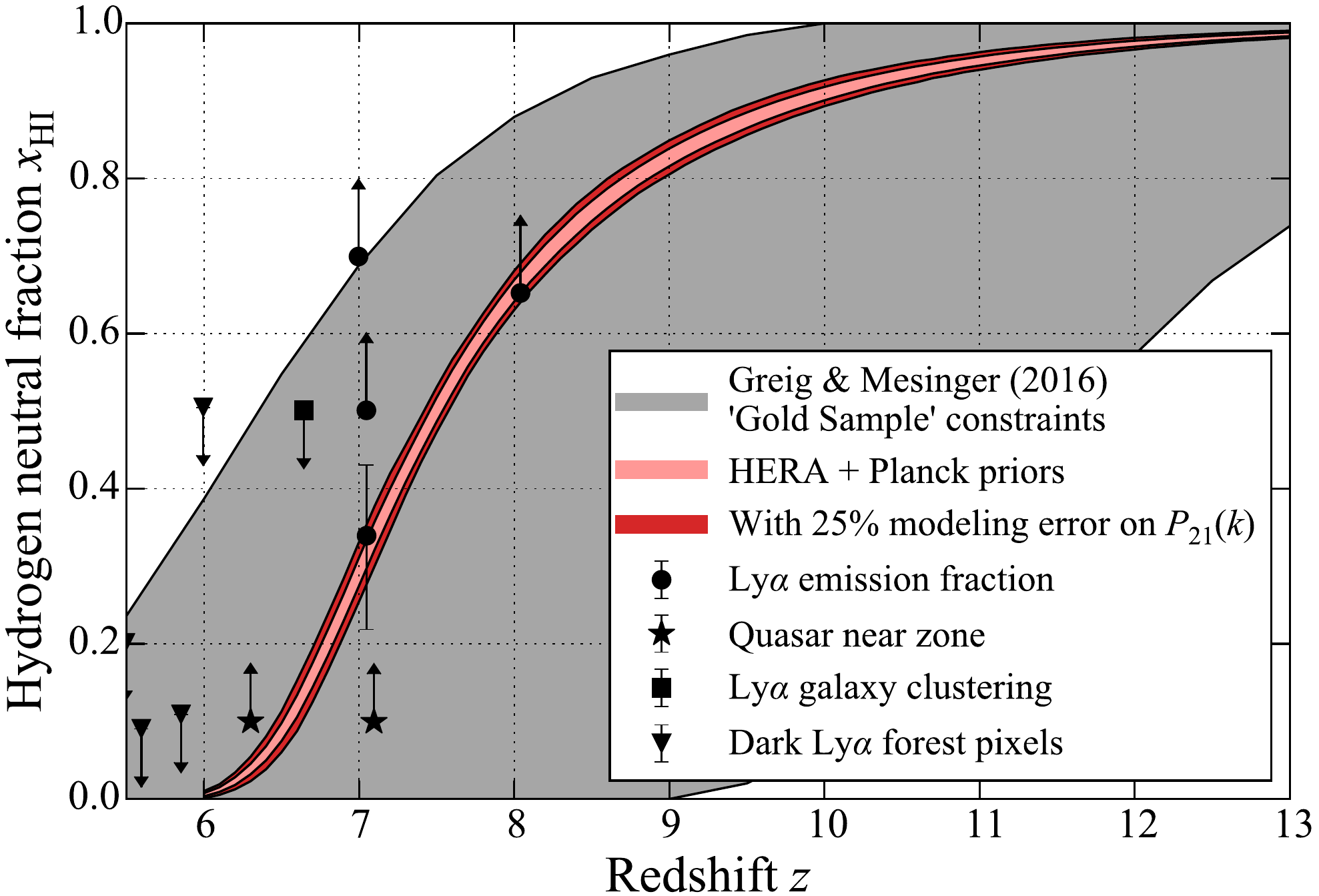}
  \caption{Combining direct constraints on $x_\text{HI}$, the hydrogen neutral fraction, as a function of redshift (black points) with \emph{Planck} priors \citep{planck_et_al2015} yields
an inferred $95\%$ confidence region (gray; Greig \& Mesinger, in prep.).  HERA constraints with (dark red) and without (pale red) 
a conservative $25\%$ modeling error in the $21\,\textrm{cm}$ power spectrum can dramatically narrow this confidence region \citep{liu_parsons2015}. Included for reference are constraints from the fraction of Lyman-alpha emitting galaxies \citep{pentericci_et_al2014,schenker_et_al2014}, quasar near-zone studies \citep{mortlock_et_al2011,bolton_et_al2011,schroeder_et_al2013}, Lyman-alpha galaxy clustering \citep{mcquinn_et_al2007_lya_clustering,ouchi_et_al2010}, and counts of dark Lyman-alpha pixels \citep{mcgreer_et_al2015}.}
	\label{fig:IonHist}
\end{figure}

Measurements of galaxy populations in
deep {\it Hubble Space Telescope} observations 
have pinned down the bright end of the galaxy luminosity function at $z \la 8$
\citep{schenker_et_al2013, bouwens_et_al2015} and are pushing deeper
(e.g.~\citealt{mcleod_et_al2015}), but producing a consistent ionization history
requires broad extrapolations to lower-mass galaxies and ad hoc assumptions about the escape fraction of
ionizing photons and the faint-end cutoff of ionizing galaxies
\citep{robertson_et_al2015, bouwens_et_al2015_reion}. 
Similarly, deducing the ionization state of the IGM from quasar proximity zones
\citep{carilli_et_al2010, bolton_et_al2011, bosman_becker2015} and the
demographics of Ly-$\alpha$ emitting galaxies \citep{fontana_et_al2010,
schenker_et_al2012, treu_et_al2012, dijkstra_et_al2014} is uncertain and
highly model-dependent.  See Figure~\ref{fig:IonHist} for these constraints on the hydrogen neutral fraction as a function of redshift.
As shown, existing probes are limited in
their ability to constrain reionization, and will be for the foreseeable future.
HERA uses
another complementary probe --- the 21\,cm ``spin-flip" transition of neutral hydrogen
--- to bring new capabilities in this area.  The next sub-sections outline these goals.

\subsection{Precision Constraints on Reionization}
\label{sec:EoRPowerSpectra}

HERA's primary science goal is to transform our understanding of the first stars, galaxies, 
and black holes, and their role in driving reionization. 
Through power-spectral measurements of the 21\,cm line of hydrogen in the primordial IGM,
HERA will be able to direct constrain the topology and evolution of reionization, 
opening a unique window into the complex astrophysical interplay between the 
first luminous objects and their environments.
The spectral nature of 21\,cm cosmology means that 
the signal at each observing frequency can be associated with an emission time (or distance) to determine both the time evolution
and three-dimensional spatial structure of ionization in the IGM.
This 3D structure encodes information about the clustering properties of galaxies,
allowing us to distinguish between models, even if they predict the same ionized fraction. 
With a new telescope optimized for 3D power-spectral measurements and with support for theoretical
modeling efforts, the HERA program will advance our understanding of early galaxy formation and cosmic reionization.

\begin{figure}
\centering
    \includegraphics[width=0.48\textwidth,clip]{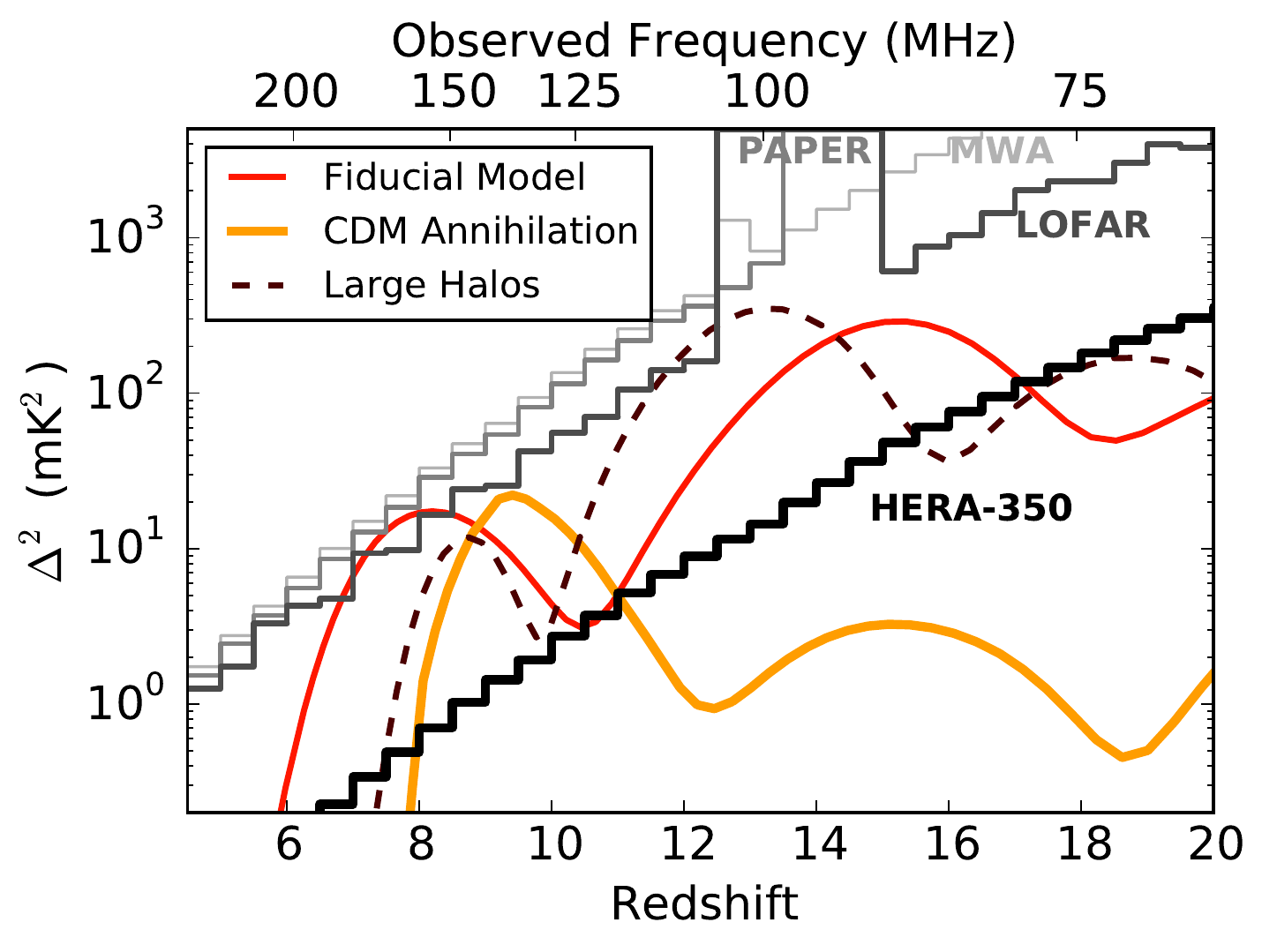}
  \caption{ 1$\sigma$ thermal noise errors on $\Delta^2(k)$, the 21\,cm power spectrum, at $k\!=\!0.2$\,$h$\,Mpc$^{-1}$ (the dominant error at that $k$)
with 1080 hours of integration (black)
compared with various heating and reionization models (colored).  Sensitivity analysis is per Table \ref{tab:signif} and associated text.}
	\label{fig:Sensitivities}
\end{figure}

HERA builds on the advances of first-generation
21\,cm EoR experiments led by HERA team members, particularly 
the Donald C. Backer Precision Array for Probing the Epoch of Reionization (PAPER; \citealt{parsons_et_al2010}),
the Murchison Widefield Array (MWA; \citealt{bowman_et_al2012,tingay_et_al2013}),
the MIT EoR experiment (MITEoR; \citealt{zheng_et_al2014}) and the Experiment to Detect the Global EoR Step (EDGES; \citealt{bowman_rogers2010}).
Recent measurements
have produced the first astrophysically constraining upper limits on the 21\,cm EoR power spectrum, 
providing evidence for significant heating in the IGM prior to reionization \citep{parsons_etal2014, 2016ApJ...820...51P, ali_et_al2015,pober_et_al2015}.
However, current experiments cannot expect more than marginal detections of the EoR signal. 
Figure~\ref{fig:Sensitivities} compares telescope sensitivities as a function of redshift to models of 
the evolving, spherically averaged 21\,cm EoR power spectrum, characterized by the dimensionless power spectrum parameter $\Delta^2 (k) \equiv k^3 P(k) / 2 \pi^2$.
The expected performance of HERA relative to current and planned telescopes to detect the peak of reionization (as well as the total collecting area) is shown in 
Table~\ref{tab:signif}.  

\begin{table}
\caption{\small Predicted SNRs of 21\,cm experiments for an EoR model with 50\% ionization at $z=9.5$, with 1080 hours observation, integrated over a $\Delta z$ of $0.8$.}
\small
 \begin{tabular}{c||r||r|r} 
\hline
Instrument & \shortstack{Collecting \\ Area (m$^2$)} & \shortstack{Foreground \\Avoidance} & \shortstack{Foreground \\Modeling} \\
\hline
PAPER & 1,188 & 0.77$\sigma$ & 3.04$\sigma$ \\
MWA & 3,584 & 0.31$\sigma$ & 1.63$\sigma$ \\
LOFAR NL Core & 35,762 & 0.38$\sigma$ & 5.36$\sigma$ \\
\textbf{HERA-350} & \textbf{53,878} & \textbf{23.34$\sigma$} & \textbf{90.97}$\boldsymbol{\sigma}$ \\
SKA1 Low Core & 416,595 & 13.4$\sigma$ & 109.90$\sigma$
\end{tabular}
\label{tab:signif}
\end{table}

These sensitivity calculations done were performed with {\tt 21cmSense}\footnote{www.github.com/jpober/ 21cmSense}\ \citep{2013AJ....145...65P,2014ApJ...782...66P}.
Foreground avoidance represents an analysis comparable to \cite{ali_et_al2015}, whereas foreground modeling allows significantly more $k$ modes of the cosmological signal to be recovered.
For the foreground avoidance approach, several design optimizations allow HERA to achieve significantly higher sensitivities than LOFAR
and comparable sensitivities to SKA, despite its modest collecting area.  The primary driver
is HERA's compact configuration.  The 21\,cm signal is a diffuse background, with most of its
power concentrated on large scales; therefore, most of an instrument's sensitivity to the EOR 
comes from short baselines.  Since HERA is a filled aperture out to $\sim 300$\,m,
for a fixed collecting area, one fundamentally cannot build an array with more short baselines
(without using smaller elements --- and HERA's dishes are already significantly 
smaller than either LOFAR or SKA stations).  
Within a $\sim150\,m$ radius from the center, LOFAR has only 11 stations,
amounting to just over $8000\,{\rm m}^2$ of collecting.  Within this radius, the SKA
is nearly filled, with $\sim 80\%$ the collecting area of HERA; 
however, the SKA underperforms in the foreground avoidance schema, where long baselines
lose more modes of the power spectrum to foreground contamination
\citep{parsons_et_al2012a}.

\begin{figure}
\centering
    \includegraphics[width=0.48\textwidth,clip]{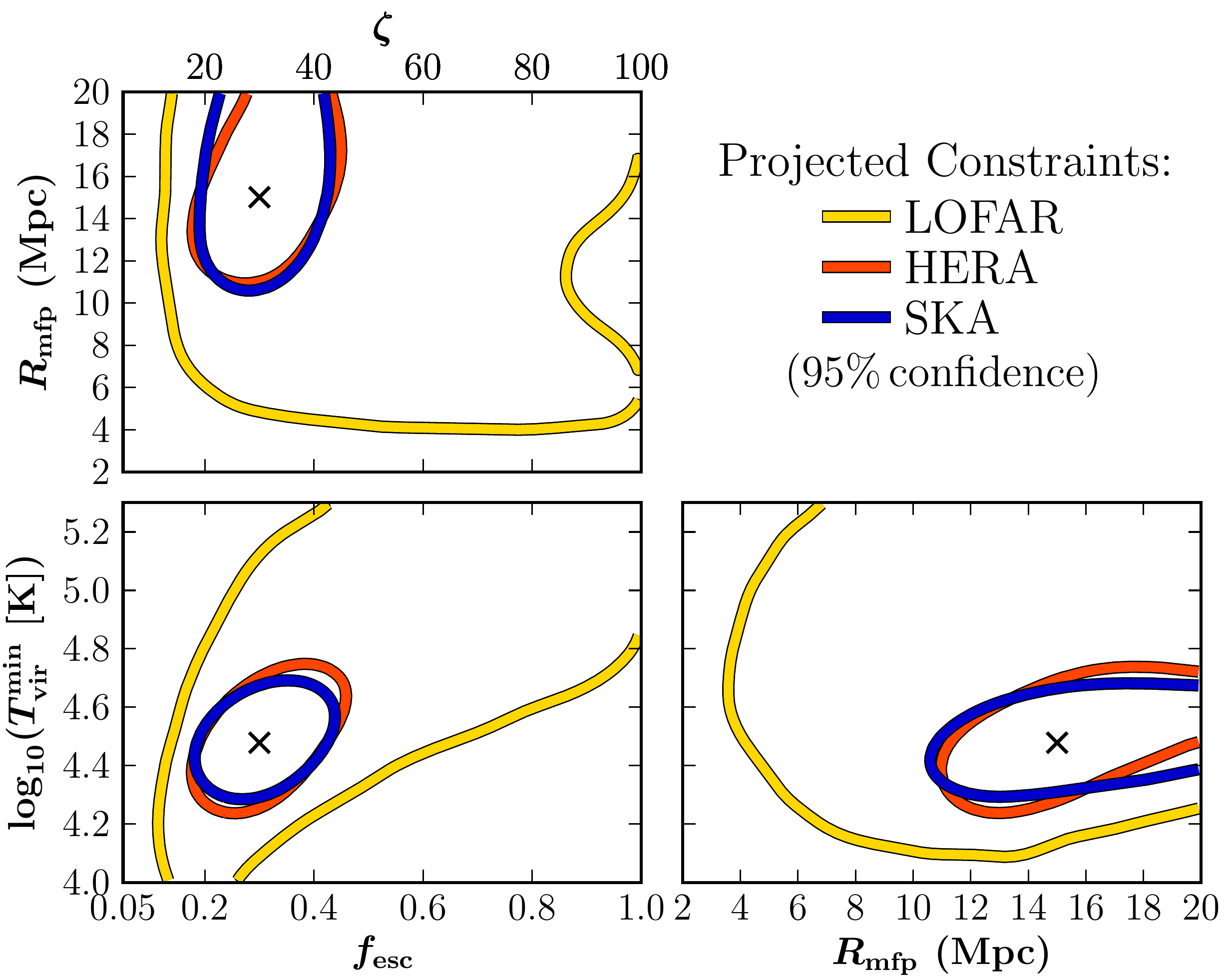}
  \caption{Projected likelihood contours from an MCMC analysis for astrophysical parameters of reionization. Model parameters are $T_\textrm{vir}^\textrm{min}$ (minimum virial temperature of ionizing galaxies); $R_\textrm{mfp}$ (mean free path of ionizing photons in HII regions); and $\zeta_0$ (ionizing efficiency of galaxies).  Also shown are constraints on the derived ionizing escape fraction, $f_\textrm{esc}$.  Adapted from
  \cite{greig_mesinger2015}. }
	\label{fig:paramConstraints}
\end{figure}
 
HERA's 21\,cm measurements can be used in conjunction with semi-analytic models to constrain the ionization history. 
The red band in Figure~\ref{fig:IonHist} shows the forecasted 95\% confidence region derived from HERA data after marginalizing over astrophysical and cosmological parameters.
Figure~\ref{fig:paramConstraints} shows the results of a Markov Chain Monte
Carlo (MCMC) pipeline for fitting models to 21\,cm power spectrum data \citep{greig_mesinger2015}, which we have conservatively limited to the range $8 < z < 10$ (although in practice a much broader bandwidth will be available; see \S\ref{sec:freqs}).
Based on the excursion-set formalism of
\citet{furlanetto_et_al2004} and the 21cmFAST code \citep{mesinger_et_al2011},
this code models the astrophysics of
reionization with three free parameters (see Fig.~\ref{fig:paramConstraints} for details). 
While the existing experiment with the most collecting area, 
the LOw Frequency ARray (LOFAR; \citealt{2013A&A...556A...2V, yatawatta_et_al2013}),
provides some ability to constrain these parameters,
HERA's constraints are significantly more precise and are comparable
to what could be achieved with the SKA. 
Additionally, HERA's constraints enable principal component parameterizations of the
sky-averaged $21\,\textrm{cm}$ signal measurements pursued by experiments such as EDGES, increasing their signal-to-noise and thus their science return \citep{liu_parsons2015}.

\subsection{Secondary Scientific Objectives}

\begin{figure}
\centering
    \includegraphics[width=0.48\textwidth,clip]{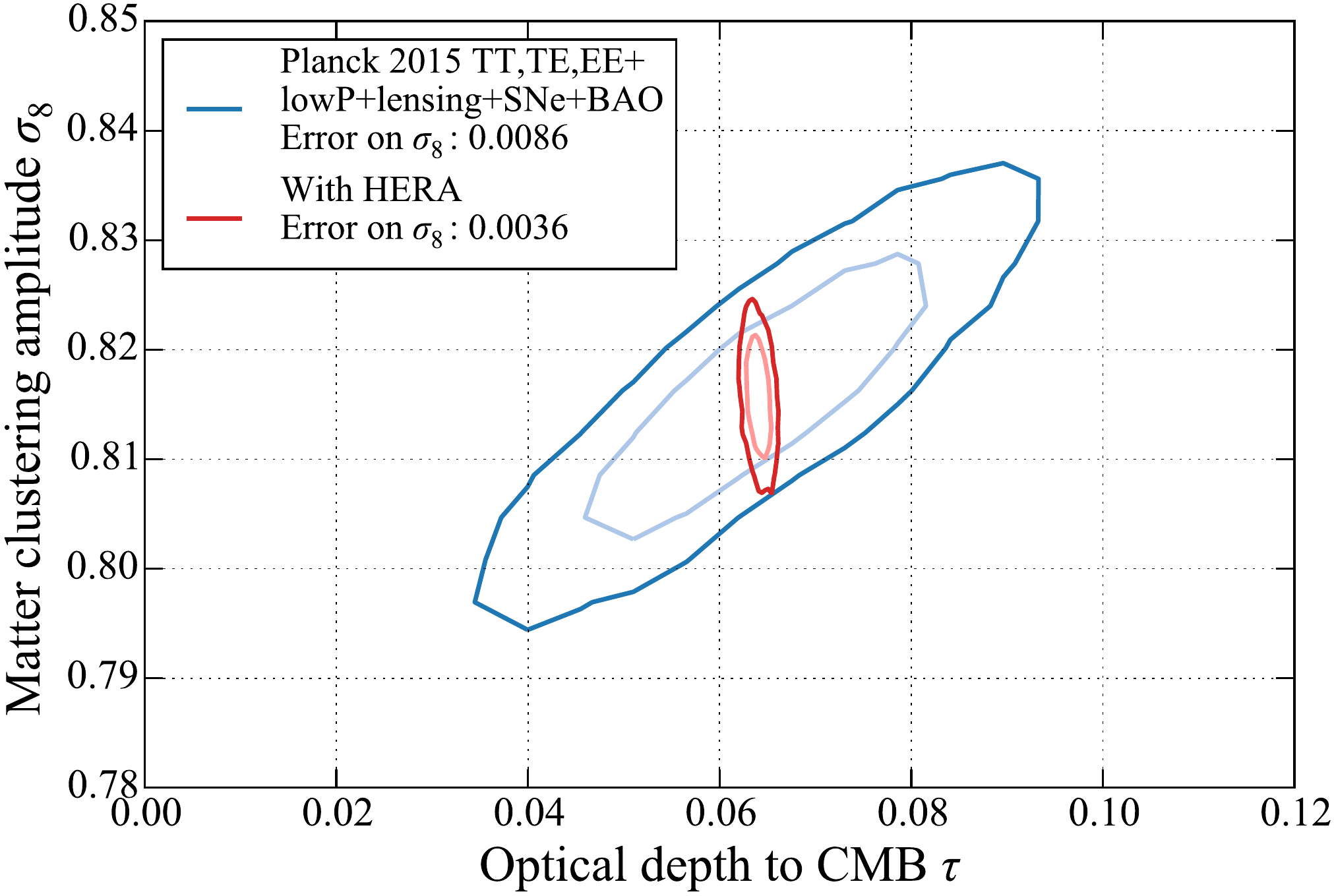}
\caption{Likelihood contours ($68\%$ and $95\%$) for $\sigma_8$ and $\tau$ using \emph{Planck} constraints (blue) and 
adding HERA data (red). 
The $21\,\textrm{cm}$ constraints break the CMB degeneracy between the amplitude of density fluctuations and the optical depth, improving constraints on both.}
	\label{fig:sigma8Tau}
\end{figure}

\emph{\textbf{Precision Cosmology.}}
\label{sec:tau}
By advancing our understanding of reionization astrophysics, HERA will improve CMB constraints on 
fundamental cosmological parameters by
removing the optical depth
$\tau$ as a ``nuisance'' parameter. HERA measurements will be able to break the degeneracy between the
constraints on $\tau$ and the sum of the neutrino masses $\sum m_\nu$, which has
been identified as a potential problem for Stage 4 CMB lensing experiments \citep{2015PhRvD..92l3535A,2016PhRvD..93f3009M}. 
A HERA-informed estimate of $\tau$ enables CMB lensing experiments to achieve a
$0.012\,\textrm{eV}$ error on $\sum m_\nu$ \citep{liu_et_al2015}. This would
represent a $\sim$5$\sigma$ cosmological detection of the neutrino masses even
under the most pessimistic assumptions
still allowed by neutrino oscillation experiments
\citep{allison_et_al2015}, making HERA key to understanding neutrino
physics. HERA's estimate of $\tau$ would also break the degeneracy between
$\tau$ and the amplitude of matter fluctuations (expressed in Fig.
\ref{fig:sigma8Tau} as $\sigma_8$) that arises when using only
CMB data. HERA effectively reduces error bars on $\sigma_8$ by more
than a factor of three \citep{liu_et_al2015}, potentially elucidating
current tensions between cluster cosmology constraints and those from primary
CMB anisotropies.

\begin{figure*}
\centering
    \includegraphics[width=1.0\textwidth,clip]{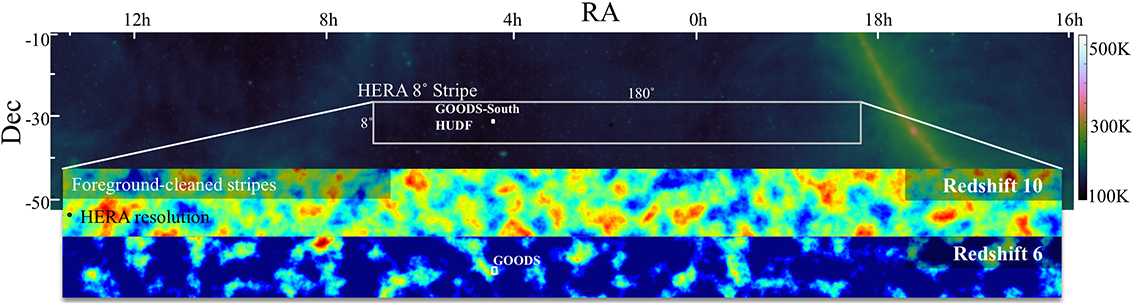}
\caption{\footnotesize 
HERA will observe a 1440 deg$^2$ stripe centered near $\delta = -30^\circ$. HERA can measure the ionization state around galaxies in, e.g., the GOODS-South field that contains a third of all known $z\!>\!8$ galaxies. HERA's primary imaging data product to the community will be deep cubes along the HERA stripe suitable for cross-correlation.
\label{fig:LSS} }
\end{figure*}

\emph{\textbf{First Images of the Reionization Epoch.}}
\label{sec:imaging}
In addition to measuring the power spectrum, there is the potential for HERA  
 to directly image the IGM during reionization
over the 1440 deg$^2$ stripe that transits overhead, which is comparable to future WFIRST
large area near-IR surveys.
 After 100 hours on a single field (achievable in 200 nights), HERA reaches a surface brightness 
sensitivity of 50 $\mu$Jy/beam (synthesized beam FWHM $\sim$ 24$'$) compared 
to the brightness temperature fluctuations of up to 400 $\mu$Jy/beam in typical 
reionization models (see Fig.~\ref{fig:LSS}). From the standpoint of sensitivity alone, HERA is capable of 
detecting the brightest structures at $z=8$ with SNR $>$ 10. Additionally, the design 
of HERA places it in a unique position to directly explore calibration techniques
(e.g. redundant calibration, \citealt{liu_et_al2010,zheng_et_al2014}), while retaining a high quality 
point spread functions for imaging and identifying foregrounds. Additional details may be found in \cite{carilli_imaging_memo}.

\emph{\textbf{Pre-Reionization Heating.}}
\label{sec:EoX} 
Prior to reionization, the 21\,cm signal is a sensitive probe of the first
luminous sources and IGM heating mechanisms. First stars are expected to form
at $z \sim 25-30$ and their imprint on the 21\,cm signal is expected to be
sensitive to the halo mass where they are formed \citep{mesinger_et_al2015}.
The IGM is then expected to be heated by first
generation X--ray binaries
\citep{furlanetto_et_al2006_global,pritchard_et_al2007,mesinger_et_al2013} or
by the hot interstellar medium produced by the first supernovae
\citep{pacucci_et_al2014}, although the heating timing and magnitude is still
very much debated \citep{fialkov_et_al2012}. Dark matter annihilation
\citep{evoli_et_al2014} could also inject energy in the IGM, leaving its
imprint in the 21\,cm signal.
 
While the observational and analytical state of the art of pre-reionization 21\,cm science is not as advanced as for the EoR \citep{2016MNRAS.460.4320E}, 
with the development of feeds sensitive down to 50\,MHz, HERA will explore the IGM prior to
reionization with a goal to constrain the sources of heating, obtaining, in the case of
X--ray heating, percent-level constraints on the efficiency with which star-forming baryons
produce X-rays \citep{ewall-wice_et_al2016a}. Lower
frequency observations also test feedback
mechanisms that interact with low-mass halos at high redshifts
\citep{Iliev_et_al2007,Iliev_et_al2012,ahn_et_al2012}.
Such constraints, while interesting in their own right, also reduce
the susceptibility of the aforementioned $21\,\textrm{cm}$-derived $\tau$
constraints to uncertainties in high-redshift physics \citep{liu_et_al2015}, especially if they are 
combined with upcoming or proposed measurements of 
the pre-reionization sky-averaged spectrum \citep{2016ApJ...821...59F}. Additionally, they may be crucial to a correct
interpretation of kinetic Sunyaev-Zel'dovich effect constraints on reionization
from the CMB \citep{park_et_al2013}. The high redshift probe of
structure afforded by the low frequency $21\,\textrm{cm}$ measurements will
permit some of the most direct observations of hypothesized suppressions of
small-scale structure
\citep{dalal_et_al2010,tseliakhovich_et_al2011,fialkov_et_al2012} arising from
predicted supersonic relative velocities between dark matter and baryonic gas
\citep{tseliakhovich_and_hirata2010}. No other electromagnetic probe can provide direct observations of this
epoch.

\subsection{Other Scientific Objectives}
\label{subsec:broader_science}

Given its high sensitivity, frequency support, filled aperture and outriggers, HERA is capable of delivering more than the core 21\,cm science discussed above.
HERA will make the data available to the community after only a short quality assurance time period to help facilitate low-frequency science.
Additionally, we encourage groups to leverage HERA as a platform upon which other equipments could be deployed. 
The Moore Foundation's support of HERA feed development and data analysis targeting the pre-reionization science described in \S\ref{sec:EoX}
is a prime example of how we envision this functioning.  Transients also provide an avenue for promising research.
Below, we list examples of the broader science that HERA could impact.

\emph{\textbf{Cross-Correlations with Other Reionization Probes.}}
\label{sec:crossCorr}
HERA's public data provide new opportunities for cross-correlation studies.
Cross-correlation between HERA 21\,cm images and other high-redshift probes
(e.g. JWST; WFIRST; CMB maps; CO, CII, and Ly-$\alpha$ intensity mapping) 
can provide an independent confirmation of the 21\,cm power spectrum
\citep[i.e.][]{lidz_et_al2009,dore_et_al2014,silva_et_al2015,vrbanec2016} and enable rich new studies of
the interaction between galaxies and their ionization environment. 
In particular, cross-correlating 21\,cm with galaxy surveys can measure the characteristic bubble size around galaxies of
different luminosities \citep{lidz_et_al2009} and help separate the degeneracy
between the fraction of photons escaping the galaxies and the
total number of ionizing photons produced \citep{zackrisson2013}.

Fortuitiously, the GOODS-South field---one of the most panchromatically studied regions of the sky, the site of the Hubble UDF, and home to over a third of all known $z>9$
galaxies---lies in HERA's field of view. HERA's images of the IGM can provide environmental context to galaxy surveys through identification of ionized bubbles ({\em e.g.} \citealt{malloy_lidz2013}), or in a more statistical sense as described in \cite{beardsley_et_al2015}.

\textbf{\emph{Searching for Exoplanetary Radio Bursts.}}
\label{sec:exoplanets}
HERA could revolutionize the study of extrasolar magnetospheres if it detects bright, highly polarized exoplanetary auroral radio bursts \citep{zarka2011}. Observers have tried to detect these distinctive events since even before the discovery of the first exoplanet \citep{winglee_et_al1986} but have thus far not succeeded. The bursts occur at the electron cyclotron frequency and are highly beamed, allowing remote sensing of planetary rotation periods and magnetic field strengths, two quantities that are otherwise only indirectly measurable. Since magnetic fields may play a key role in protecting atmospheres and/or biospheres from energetic stellar wind particles \citep[e.g.,][]{tarter_et_al2007}, studies of planetary magnetism may be vital to evaluating habitability. HERA's sensitivity, large field of view, long campaigns, and precise calibration make it well-suited to the search for radio bursts from Jupiter-like planets out to 25~pc. Well-studied exoplanet host stars in the HERA stripe within that distance include Fomalhaut, Gl~317, Gl~433, Gl~667~C, and HD~147513.

\textbf{\emph{Fast Radio Burst Followup.}}
\label{sec:FRBs}
Fast Radio Bursts (FRB) are millisecond-long radio flashes whose origin has remained a great enigma ever since their discovery \citep{2007Sci...318..777L}. 
HERA could be triggered by nearby, higher-frequency telescopes for FRB followup, saving baseband data and thus full sensitivity to all dispersion measures.
This would require additional hardware, but the architecture allows for such additions.
Bursts like that discovered by \citet{masui_et_al2015}, the lowest frequency
FRB detection to date, should be seen hourly by HERA at 5--10$\sigma$.
 Observations at HERA frequencies are very sensitive to the physics of
the intervening medium, particularly deviations from $\lambda^2$
dispersion. Detecting deviations would rule out broad classes of models and could indicate whether FRBs are at cosmological distances.

\begin{figure}[h!]
\centering
\vspace{0pt}
    \includegraphics[width=0.48\textwidth,clip]{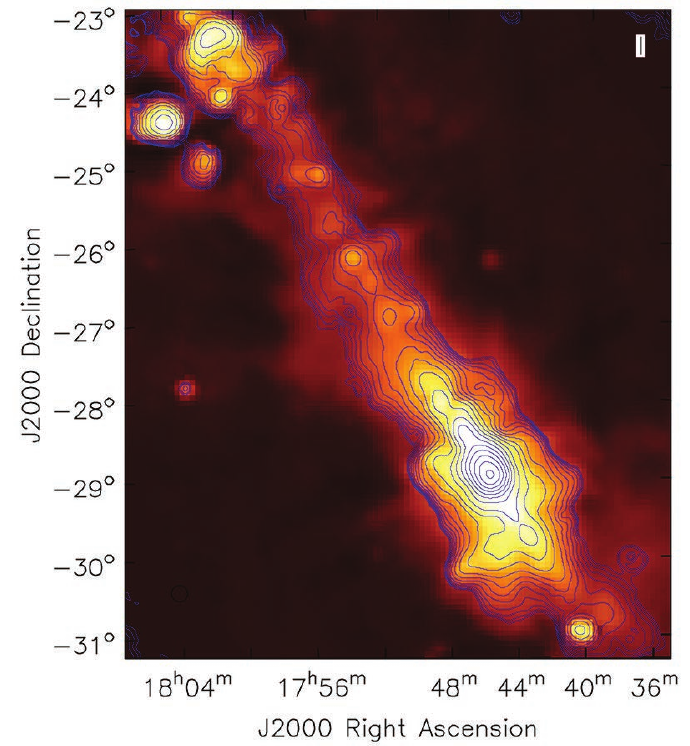}
  \vspace{0pt}
\caption{\footnotesize 
The color-scale shows an input model image of the
Galactic center region based on the Effelsberg 21cm continuum survey
\citep{1990A&AS...83..539R}.  The
contours show the mock HERA-350 observation at 130MHz with a resolution of
$12'$. Contour levels are a geometric progression in square root 2,
starting at 5 Jy beam$^{-1}$. The peak surface brightness on the HERA
image is 2330 Jy beam$^{-1}$.
\label{fig:galcen} }
\vspace{0pt}
\end{figure}

\textbf{\emph{Continuum imaging.}}
Figure \ref{fig:galcen} shows a mock observation of the Galactic Center region.
For the input
model sky (color scale), we adopt the Effelsberg 1.4GHz sky survey at $\sim 10'$
resolution and a pixel size of 4$'$ \citep{1990A&AS...83..539R}.  The model image
is adjusted to Jy pixel$^{-1}$, and scaled to 130MHz assuming a
spectral index of $-0.8$.   Noise is added
per visibility assuming standard array parameters, and an integration
time of 100hrs, and a bandwidth of 8MHz. The thermal noise in the
image in bright regions of the Galaxy would be about 100$\mu$Jy,
however, at $12'$ resolution and 130MHz the array is completely
in-beam confusion limited at the level of a few Jy beam$^{-1}$.
The resulting image is shown as the blue contours.  See \cite{carilli_imaging_memo} for more details.

\section{Measuring the EOR}
\label{sec:eormeas}
Due to the expansion of the universe over cosmological time, we can identify and measure the early Universe via the redshift of spectral lines.  The hydrogen hyperfine transition at a rest frequency of 1420 MHz is a key spectral line due to the ubiquity of hydrogen and, being a ``forbidden'' transition, the optical depth lets us see through the entire universe back almost to the period of recombination.  The bandwidth
therefore equates to a cosmic distance along the line-of-sight of the telescope, with the frequency determining the cosmic age.  An observation of an area of sky over a given bandwidth is therefore providing an average
over a cosmic volume.

As the distance to the EOR is great (about 13 billion light-years) the signal is weak.  However, the signal also subtends the entire sky, so initial measurements of the EOR strive to measure a statistical power spectrum of the signal over the sky since the nature of the reionization process should have a specific spatial signature.  The goal is therefore to measure a range of aggregate spatial scales on the sky, rather than to image the signal directly.  Imaging does remain an ultimate goal to fully understand the process, however we will likely need a greater understanding of the signal characteristics and the systematics to achieve this more difficult goal.  

Obviously between our present observation point and the Epoch of Reionization lies the entire intervening universe, which has an intrinsically much brighter signal (up to 6 orders of magnitude or more).
Primarily, this power is due to diffuse Galactic synchrotron radiation, supernova remnants and extragalactic radio sources.  As a first step, areas of the sky where these signals are minimal (for example, outside the galactic plane and away from strong point sources) are targeted \citep{bernardi_2013}.  More importantly and most fortuitously however is the fact that all of these foreground signals are smooth spectrum sources whereas the expected spectrum of the EOR is expected to be rough since it is made up of non-ionized regions which are randomly distributed over a wide range of redshifts.  This fact allows us to try and isolate foregrounds from the EOR, as will be discussed.

\begin{figure}
	\centering
	\includegraphics[width=.48\textwidth]{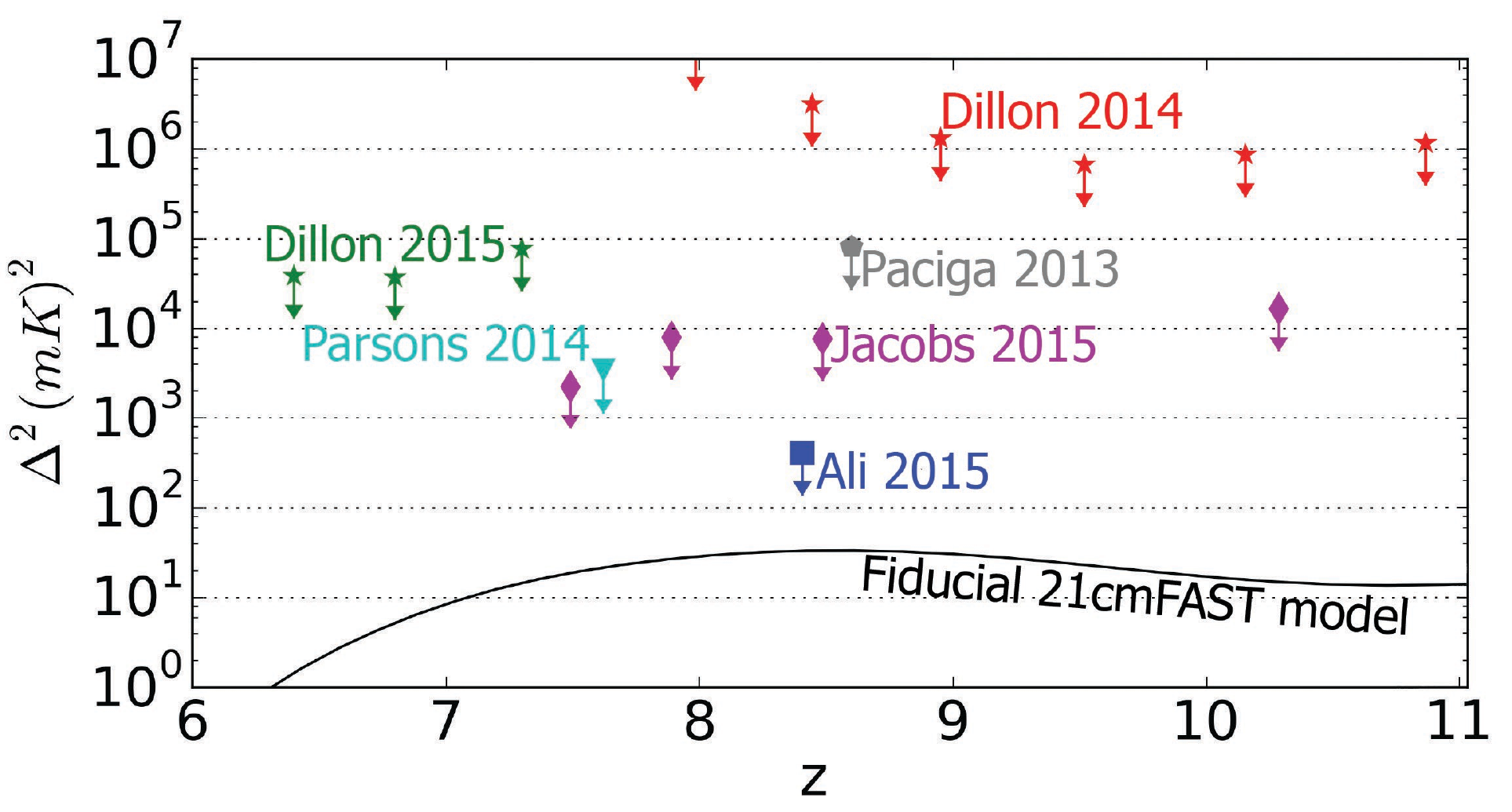}
	\caption{The current best published $2\sigma$ upper limits on the 21cm power spectrum, $\Delta^2(k)$, compared to a 21cmFAST-generated model at $k=0.2$\,$h$\,Mpc$^{-1}$.   Analysis is still underway on PAPER and MWA observations that approach their projected full sensitivities;  HERA can deliver sub-$\text{mK}^2$ sensitivities.}
	\label{fig:limits}
	\label{fig:IGMtemperatureConstraints}
\end{figure}

As seen in Figure~\ref{fig:limits}, the past three years have seen deep EOR observations with PAPER, the MWA, 
the Giant Metrewave Radio Telescope (GMRT), and LOFAR.
PAPER and the MWA have produced progressively deeper limits
\citep{dillon_et_al2014,dillon_et_al2015,parsons_etal2014,ali_et_al2015}, with PAPER
yielding the first meaningful constraints on the 21\,cm spin temperature during reionization.
The inherent challenge is to simultaneously meet stringent sensitivity requirements 
while suppressing foregrounds $\sim5$ orders of magnitude
brighter than the 21\,cm signal, which has been addressed along a number of approaches.  
To address this challenge,  all aspects of the experimental process must be refined and improved, spanning 
calibration \citep[{\em e.g.},][]{zheng_et_al2014, 2016ApJ...825..114J, 2016MNRAS.461.3135B}, handling foreground contamination
\citep[{\em e.g.},][]{moore_et_al2013,thyagarajan_et_al2015a,thyagarajan_et_al2015b,2016ApJ...819....8P},
and the interferometer design itself \citep[{\em e.g.},][]{parsons_et_al2012a,dillon_parsons2016}.

A promising approach to make the first detection of the EoR power spectrum 
is a foreground mitigation strategy based largely
on identifying and filtering out a region of parameter space where the strong foregrounds are largely confined (\S\ref{sec:delayapproach}).  This is the approach
pioneered by PAPER.
By reducing the need
for foreground modeling and subtraction, this approach allowed PAPER to switch to a grid-based antenna layout that
enhanced sensitivity toward fewer Fourier modes and facilitated calibration
based on the ``redundancy" of different antenna pairs measuring the same sky modes \citep{liu_et_al2010,zheng_et_al2014}.
Combined, redundancy and delay filtering provide a robust, inexpensive, and demonstrably successful solution 
to the foreground problem.

Yet PAPER's lack of imaging support and its uneven $uv$ sampling 
leave it with limited diagnostic capability, particularly for direction-dependent systematics
such as polarization leakage from Faraday-rotated emission \Mycitep{moore_et_al2013, nunhokee_etal_2016}.
While concern over such effects has decreased markedly since discovering that
intrinsic point source polarization is lower than previously thought
\citep[]{bernardi_2013,asad_et_al2015} and that variable Faraday rotation through the
ionosphere averages down the polarized signal over long observing seasons
\Mycitep{moore_et_al2016,aguirre_submitted_2016}, direction-dependent beam effects remain an area of interest.
The MWA's image-based calibration and foreground subtraction strategy provides complementary
capabilities (\S\ref{sec:mapapproach}). Imaging with subtraction, while still under development as a viable foreground strategy,
can increase sensitivity by recovering more modes of the cosmological signal (see
Table~\ref{tab:signif}) and help address systematic errors rooted
in the image domain.  HERA's antenna configuration---shown in Figure~\ref{fig:arrayConfig} and discussed in \S\ref{sec:arrayConfig}---emphasizes
the proven approaches of redundant calibration and delay filtering, while simultaneously
increasing the extent and density of $uv$ sampling for high-fidelity imaging.  

Despite progress, the fact remains that the 21\,cm EoR signal is intrinsically very faint; making a detection requires a
large instrumental collecting area and a long, dedicated observing campaign.
Although published PAPER and MWA results (Fig. \ref{fig:limits}) do not yet include observations 
at full sensitivity that are still being analyzed (e.g. Fig. \ref{fig:Sensitivities}), 
it is clear already that these instruments lack the sensitivity
to make a conclusive detection (see Table~\ref{tab:signif}). 
HERA addresses this shortcoming with a dish element that delivers a much larger collecting area 
while retaining the necessary characteristics for both proven and forward-looking foreground removal strategies.

This section will provide a brief overview of  the theoretical underpinning of how foregrounds contaminate the measurement (the so-called ``wedge''), discuss the various techniques used to make the measurement and finally provide a brief discussion on calibration issues and techniques.

\subsection{The ``Wedge''}
\label{sec:wedge}
\begin{figure*}
	\centering
	\includegraphics[width=.94\textwidth,clip]{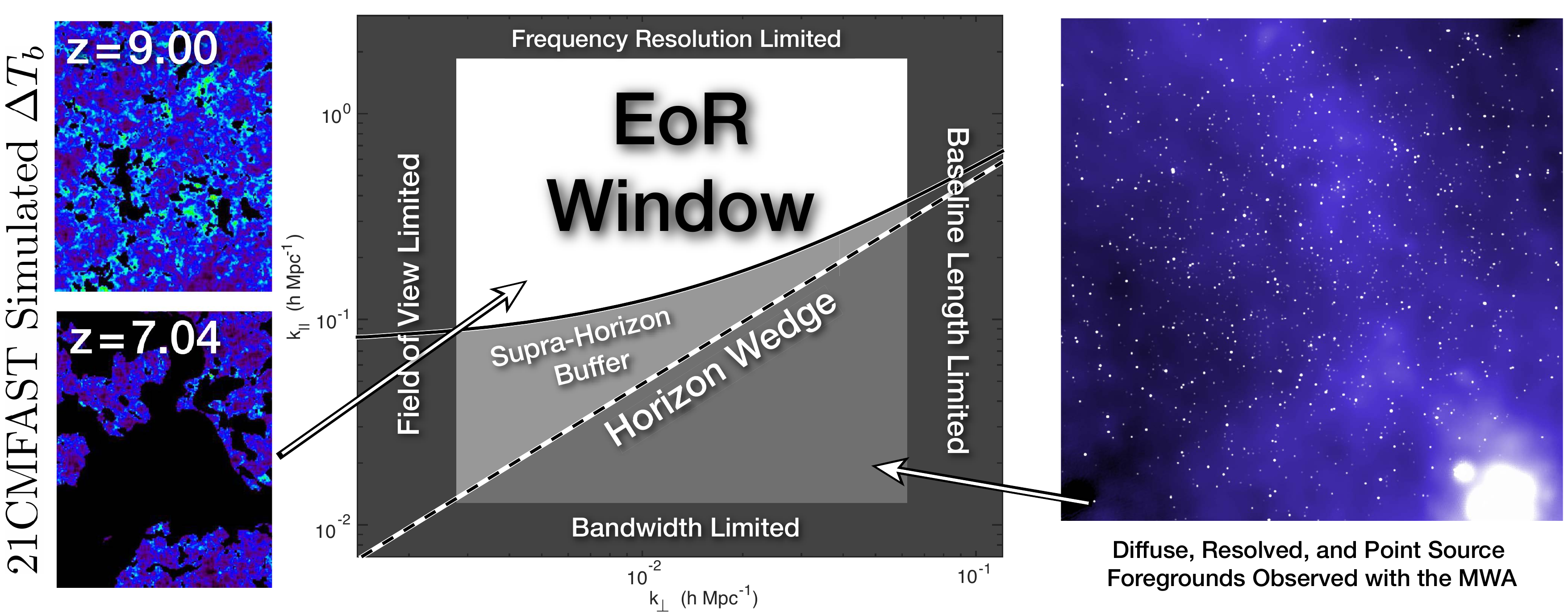}
	\vspace{-10pt}
	\caption{Foregrounds are a primary challenge facing 21\,cm cosmology experiments. 
HERA leverages a ``wedge'' in the cylindrically-averaged ($\mathbf{k}$ is broken into $k_\|$ and $k_\perp$) power spectrum (center panel). Smooth-spectrum foregrounds (right panel) $\sim 3-5$ orders of magnitude brighter than fiducial EoR models (left panel; \citealt{mesinger_et_al2011}) create the ``wedge'' when they interact with the interferometer's chromatic response. Flat spectrum foregrounds cannot contaminate the EoR window beyond the horizon wedge, though intrinsic or imperfectly calibrated chromaticity can push it slightly beyond the horizon. By avoiding foregrounds, PAPER has placed limits within an order
of magnitude (in mK) of these models \citep{ali_et_al2015} and shown the ``EoR Window'' to be foreground-free.
HERA's dish and compact configuration optimize wedge/window isolation and direct sensitivity to low-$k_\perp$ modes where the EoR is brightest relative to the foreground wedge.
}	\label{fig:wedge}
\end{figure*}

Perhaps the most important advance informing HERA's design is a
refined understanding of how smooth-spectrum foregrounds interact
with instrument chromaticity to produce a characteristic ``wedge" of
foreground leakage in Fourier space (see Fig.~\ref{fig:wedge}), 
outside of which the 21\,cm signal dominates (the ``EOR window'').  The wedge is a direct consequence of the chromatic response of an interferometer.
Through theoretical and observational work
\citep{datta_etal2010,morales_et_al2012,parsons_et_al2012b,vedantham_2012,thyagarajan_et_al2013,hazelton_et_al2013,pober_etal2013b,liu_et_al2014a,liu_et_al2014b},
we have learned how the boundary between the wedge and  the EoR window is determined by the separation between antennas,
signal reflections within antennas, and the angular response of the antenna beam.  Deep integrations also show us
that, to the limits of current sensitivity, intrinsic foreground emission is absent outside of the wedge; it can only 
appear there through instrumental leakage \citep{parsons_etal2014,ali_et_al2015,moore_et_al2016,kohn_et_al2016,2016ApJ...825....9T}.

The power spectrum measurement provides the spatial correlations across the sky, characterized by the magnitude of the wavenumber, ${\bf k}$.
Though the full magnitude of the {\bf k}-vector is used, it is instructive to split it into two components,  $\kvec =  \kvpr + \kpar{\hat {\bf z}}$ where $|\kvpr|\equiv \kpr = 2\pi b/\lambda X$ is determined by the antenna baseline ($b$) and $\kpar=2\pi/(YB)$ by the bandwidth ($B$).  Here $X$ and $Y$ are cosmological parameters relating angular size and spectral frequency to cosmic volumes respectively (so, relating wavenumber to physical volume at a given redshift).  
$\kvpr$ corresponds to the plane of the sky and $\kpar$ to the line-of-sight.
This is useful since it allows us to split the chromatic response of the interferometer visibility measurement from the instrument bandpass and isolate a phase space where smooth-spectrum foreground sources ({\em i.e.} everything {\em not} the EOR) contaminate the signal of interest from where they don't.  

As just stated, the $\kvpr$ components are directly proportional to the baselines and $\kpar$ are proportional to the Fourier transform of the frequency response \footnote{Note that this equivalency is an approximation, which is very good for the short baselines and small bandwidths generally used here.  For a more complete discussion see \cite{liu_et_al2014a,liu_et_al2014b}}.   
The Fourier transform of a frequency spectrum is a delay spectrum, hence the ``delay-spectrum'' moniker for the technique that uses this approach.  Note that cosmic evolution limits the largest bandwidth (which determines the smallest $\kpar$) to about 10 MHz -- for larger bandwidths the evolution of the Universe begins to impact the result.  For HERA, wavenumbers are dominated by the bandwidth, not the baseline.

\begin{figure}
\centerline{
\includegraphics[width=.48\textwidth]{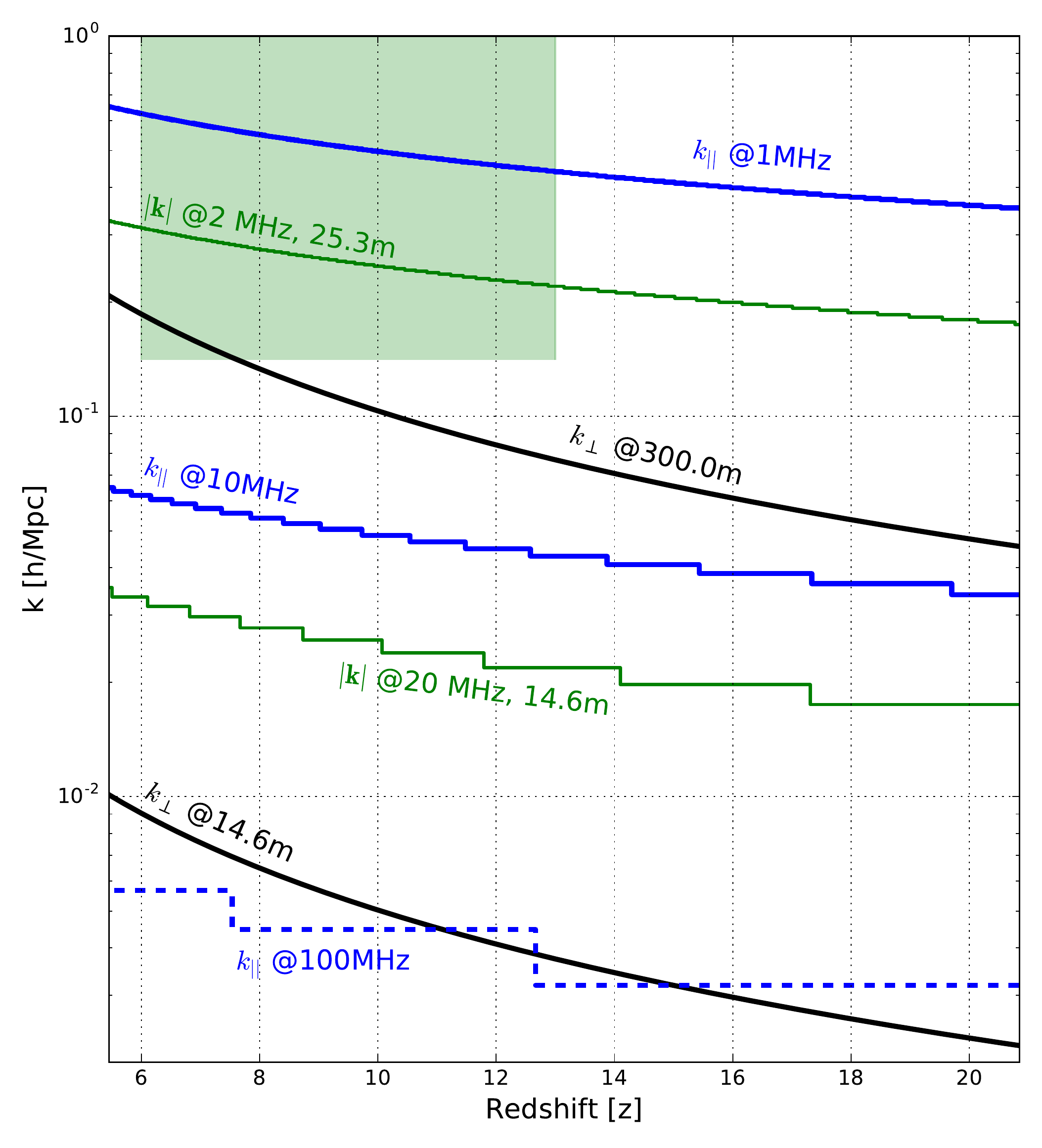}
}
\caption{\small $|\kvec |, \kpar, \kperp$ wavenumber plots at various bandwidths ($k_{||}$; blue), baselines ($\kperp$; black), and total ($|\kvec |$, green) as a function of redshift.  The stepped profiles for $k_{||}$ and $|\kvec |$ are the line-of-sight redshift ranges for that bandwidth (too small to see at 1 MHz).  See text for discussion.  The full redshift range is the HERA extended goal.  The initial HERA delay-spectrum power spectrum goal is the light green square region in the upper left.}
\label{fig:kperf}
\end{figure}

Figure \ref{fig:kperf} shows many of the dependencies on wavenumber of the bandwidth, configuration and cosmology by plotting the perpendicular wavenumber ($\kvpr$; black lines), parallel wavenumber ($\kpar$; blue lines) and total wavenumber ($k$; green lines) as a function of redshift for various bandwidths and baselines.  The redshift range is appropriate for the extended frequency range that is the goal for HERA, however $z=6-13$ is the primary region for EOR studies.
The lines are labelled for the assumed bandwidth and/or baseline, depending on the type of wavenumber shown.  Obviously many values may be used, but here a small sampling of values appropriate for HERA are shown.    
Wavenumbers with an assumed bandwidth ($k$ and $\kpar$) are shown as stepped profiles, with the redshift steps equal to the bandwidth at that redshift.
Note that for 100 MHz, the concept of a power spectrum at a given redshift is no longer appropriate, which is why the line is dashed.
Also in Figure~\ref{fig:kperf}, the lime-green shaded area indicates the initial region in which HERA intends to detect and characterize the EOR using the delay-spectrum technique.

Note that the unit of the spatial wavenumber $k$ is length$^{-1}$, in this case the relevant length scale is megaparsecs (Mpc; 1 Mpc = 3.086$\times10^{22}$ m).  In order to account for updated measurements of the present-day Hubble parameter $H_0$, it is further normalized by a factor $h$, where $H_0=100h$ (km/s)/Mpc, such that the wavenumbers are expressed in units of $h$~Mpc$^{-1}$.  Also note that at the redshifts of interest $X$ has a value of about 160 Mpc~deg$^{-1}$ and $Y$ has a value of about 16 Mpc~MHz$^{-1}$.

In these variables, the ``contaminated'' phase space is a wedge-shaped region in $\kpr - \kpar$ space such that the ratio of the time-delay across a given baseline ($b/c$,  which impacts the chromatic visibility) and the delay associated with a given bandwidth ($1/B$) is less than some parameter determined by the details of the system, which we denote as $1/\beta$.  Substituting $\kpr$ and $\kpar$ in for $b$ and $B$ in this ratio, this wedge is bounded by
\begin{equation}
\label{Eq:wedge}
\kpar \le \beta\frac{X\lambda}{Yc}\kpr + \frac{S}{Y}
\end{equation}
where an offset parameterized by $S$ accounting for effects related to the combined spectral smoothness of the foregrounds and the antenna response has been included (see Fig. \ref{fig:wedge}). The impact of this offset parameter is explored in \citet{thyagarajan_et_al2013}.

If this $\beta$ factor is determined solely by the chromaticity of the longest baseline, we see that $\beta=1$ and this line is referred to as the ``horizon line'' ({\em i.e.}, the delay of a source at the horizon).  Systematics will push that line such as to close the EOR ``window'' to the upper left ($\beta > 1$).
If we can completely control systematics and constrain the effects of the foreground to {\em e.g.} the main beam, the line will move to decrease the size of the wedge ($\beta<1$).
Therefore, a key issue to measuring the EOR power spectrum is to understand and minimize systematics.  For further discussion and an analytical derivation, see \citet{vedantham_2012}, \citet{thyagarajan_et_al2013}, \citet{liu_et_al2014a} and \citet{liu_et_al2014b}. In simulations see \citet{datta_etal2010} and \citet{hazelton_et_al2013}, and for observations
\citet{pober_etal2013b}, and \citet{parsons_etal2014}.  

The techniques to measure the power spectrum may be broken down into two principle techniques -- delay-space (\S\ref{sec:delayapproach}) and map-making (\S\ref{sec:mapapproach}) -- plus additional hybrid methods.  These are briefly summarized below, with an emphasis on the delay-spectrum technique as the initial approach taken by HERA.  Finally, we provide a short discussion on calibration in the concept of this design.

\subsection{Delay-Spectrum Approach}
\label{sec:delayapproach}
Given that the response of an interferometer natively measures the power in Fourier modes of the sky within its beam, we see that it is a natural instrument to use for this measurement of the EOR spatial power spectrum.  The delay-spectrum approach leverages the interferometer measurement to optimize sensitivity to the desired modes while rejecting modes contaminated by the foreground power in the wedge.  Other than potentially handling overlapping bins in {\em uv} space, the delay-spectrum approach does not combine baselines before squaring and calculating the power spectrum, which contrasts to the map-making techniques briefly described in the following section.  For more details, see \cite{parsons_et_al2012b}.

We can approximate the sky power spectrum $P(\kvec)$ as being linearly proportional to the Fourier transform along the frequency axis (the delay-transform) of an interferometer baseline visibility, denoted ${\tilde V}_b(\tau)$:
\begin{equation}
P(\kvec) \approx\frac{X^2Y}{4k_B^2}   \left[\frac{{\tilde V}^2_b(\tau)}{\Omega_b B/ \lambda^4} \right]
\label{eq:Pk_baseline}
\end{equation}
where $X$ and $Y$ were introduced above, $\Omega_b$ is the integrated beam response, $B$ is the effective bandwidth, $\lambda$ is the observation wavelength, and $k_B$ is Boltzmann's constant.  The terms in square brackets are instrumental terms, as opposed to the constants and cosmological parameters out front.  
Rather than $P(\kvec)$, the literature generally works with a volume-normalized parameter given by $\Delta^2(k) = \frac{k^3}{2\pi^2}P(\kvec)$.    

Since the thermal noise per visibility baseline may be expressed as
\begin{equation}
V_N = \left(\frac{2k_B}{\lambda^2}\right)\left(\frac{T_{sys}}{\sqrt{2Bt}}\right)\Omega_b B
\label{eq:sensitivity_per_baseline}
\end{equation}
where $t$ is the integration time (see {\em e.g.} \citealt{thompson_et_al2001}),
to determine the sensitivity of an instrument to the power spectrum per baseline we can substitute the thermal noise per baseline for the visibility in Eq. \ref{eq:Pk_baseline}.  Further, since an interferometer typically measures many baselines which may be averaged together to improve the signal-to-noise per $k$-bin, the total sensitivity may be approximated as
\begin{equation}
\Delta^2_N (k,z)\approx \frac{k^3X^2Y}{4\pi^2} \left[\frac{T_{sys}^2\Omega_b }{t \mathcal{N}}\right]
\label{eq:sensitivity}
\end{equation}
where $\mathcal{N}$ represents the improvement in sensitivity based on the array configuration, which may significantly boost the sensitivity if optimized for this measurement (see \citealt{parsons_et_al2012b}).  

\begin{figure*}
\centerline{
\includegraphics[width=0.95\textwidth]{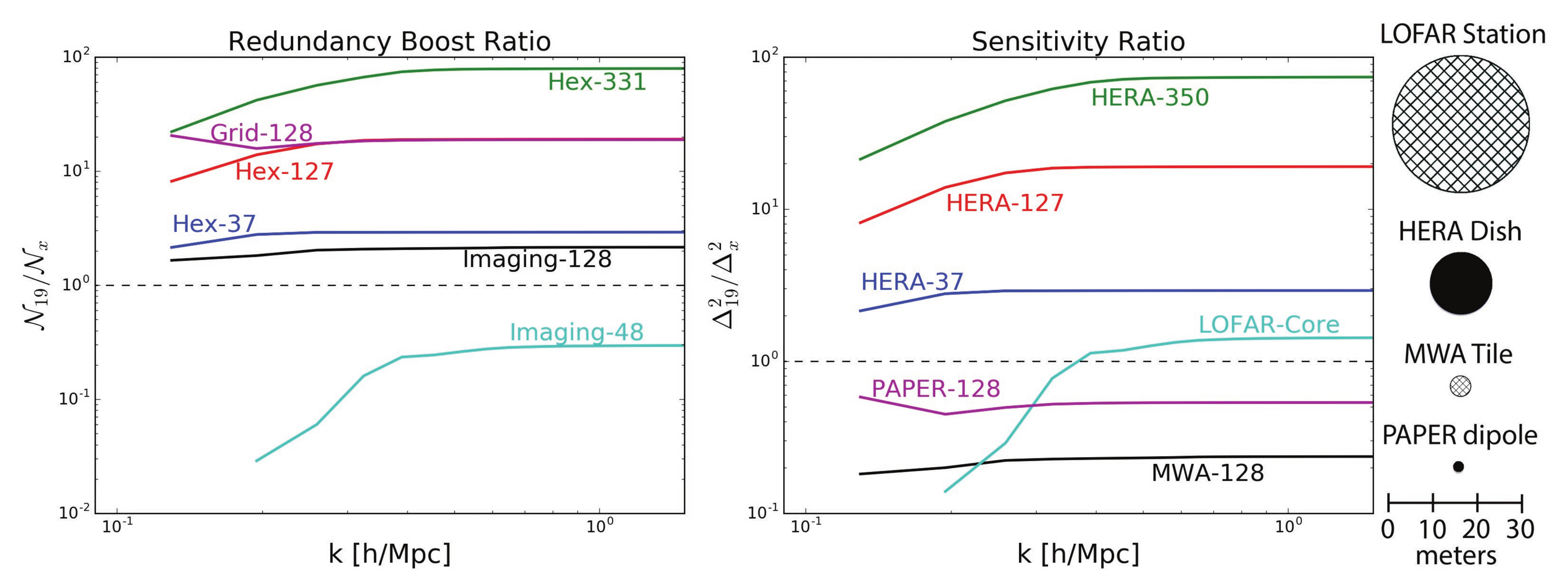}
}
\caption{Left: Representative boost factors ($\mathcal{N}$ in Eq. \ref{eq:sensitivity}) normalized to a 19-element hexagonal configuration spaced at 14.6~m.  For a given element count, one can get about an order of
magnitude improvement in delay-spectrum sensitivity.  Alternatively, for the same sensitivity one can build roughly 1/3 the antennas.  Right:  Power spectrum sensitivities for various arrays normalized to HERA-19.  The element sizes are indicated at the far right with appropriately sized circles.  Cross-hatching denotes phased tiles.}
\label{fig:boost}
\end{figure*}

Figure~\ref{fig:boost} (left) plots this ``redundancy boost'' factor for various configuration types relative to an array in a 19-element hexagonal configuration.  They are labelled for the type of configuration ({\em hex, grid, imaging}) and the number of antennas (from 37 to 350).  Note that (a) {\em hex} is based on the configuration used for HERA (14.6m center-to-center spacing); (b) {\em grid} for that of PAPER (4m N-S, 16m E-W spacing); (c) {\em imaging-128} for MWA \citep{tingay_et_al2013}; and (d) {\em imaging-48} for LOFAR (\citealt{2013A&A...556A...2V}; note that this assumes the ability to fully correlate the HBA-pairs in the Netherlands core).  The plot shows that for a fixed number of elements using the delay-spectrum approach, redundant arrays ({\em hex} and {\em grid}) provide about an order of magnitude improvement over imaging arrays.  Since the baselines go as order $N^2$, this corresponds to using about 1/3 of the number of elements to yield the same performance (as verified by the {\em Hex-37/Imaging-128} lines).  This specific dependency does not extend to the other approaches discussed below and imaging configurations would likely be preferred.  Note that a filled hexagonal-packed array provides excellent imaging as well, but with potentially limited resolution unless outriggers are included.

Another approach to improving sensitivity is to make the collecting area per element larger.  A larger element does limit the accessible field-of-view, but this is not a huge liability for a focussed experiment as long as an appropriate patch of ``cold'' sky passes within its field-of-view and a large enough piece of sky is surveyed to overcome sample variance.  These conditions hold true for HERA.  Figure~\ref{fig:boost} (right) shows the relative sensitivity incorporating configuration and element size of various arrays relative to HERA-19 using  {21cmSense}, calculated for a redshift corresponding to the peak signal at a hydrogen reionization fraction of 0.5 (in this case, z=8).  The circles to the right indicate the relative sizes of the elements.  Note that MWA and LOFAR are cross-hatched to indicate phased array tiles, which uses small antennas to phase up a full element response.

As an aside, we can compare and contrast the delay-spectrum technique to imaging with an interferometer.  For imaging, one takes a 2-D spatial Fourier transform of the visibilities, and produces images as a function of frequency.  Here, one is taking a 1-D Fourier transform along the other axis (the bandpass) and produces a 3-D spatial power spectrum.  In imaging, point sources are transformed to a single pixel, whereas for the delay-spectrum, smooth sources are transformed to a single delay.  For imaging, we are concerned with ripples in the passband due to standing waves, whereas for the delay-spectrum we are concerned with delayed versions of the signal due to multiple reflections (equivalent views of the same phenomena).

\subsection{Mapmaking Approach}
\label{sec:mapapproach}
In contrast to the delay-spectrum approach, mapmaking approaches combine baselines to build up information before squaring and calculating the power spectrum.  The conceptually simplest approach is to first make an image cube of the sky as a function of angular position and frequency, take the spatial Fourier transform, square and bin the cube, then subtract the dominant foreground power and take the transform to determine the EOR power spectrum \citep{liu_tegmark2011,dillon_et_al2013a}. The image domain is a natural place to combine information from partially-coherent pairs of visibility measurements (either from different baselines or from the same baseline at different times). Mapmaking can be a lossless data compression step \citep{tegmark1997b}, though keeping track of statistical properties of the maps, which have complex frequency and position dependent point spread functions and noise covariance matrices, can be computationally challenging \citep{dillon_et_al2015a}. Existing approaches have had to make a number of approximations, including that point spread functions do not vary over the field of view and that noise is not correlated between $uv$-cells. Additionally, mapmaking is in principle not as vulnerable to polarization leakage as the delay-spectrum approach, since only polarization mismodeling (rather than any kind of polarization asymmetry) can cause leakage from Stokes $Q$ or $U$ to $I$.

One advantage of such an approach is that sky images can be interesting in their own right, both for dealing with measurement systematics and for accessing the non-Gaussian observational signatures (such as ionized bubble structures) that are missed by variance statistics such as the power spectrum. Mapmaking also allows for the direct subtraction of bright foregrounds, potentially expanding the EoR window. However, accessing these high-order statistics or the modes of the power spectrum inside the wedge using a mapmaking approach requires extreme (and so-far undemonstrated) precision in calibration and forward-modeling of bright foreground models through instrument systematics. For this reason, the approach is to detect and initially characterize the power spectrum via the delay-spectrum technique while developing tools for other approaches.

\subsection{Hybrid Approaches}
\label{sec:hybridapproach}

While they were presented separately above, the delay-spectrum and map-making approaches need not be viewed as mutually exclusive. For example, mapmaking does not nned to be limited to the real space image basis; pipelines that combine multiple baselines into estimates of various Fourier amplitudes of the sky are formally also mapmaking pipelines. Like the delay-spectrum approach, such pipelines (see, e.g., \citealt{trott_et_al2016}) avoid the image domain entirely. This helps to prevent artifacts that may be introduced by imaging algorithms.

More generally, it is possible to express a mapmaking algorithm as a linear operator that acts on visibility data to produce a compressed dataset \citep{dillon_et_al2015a}. It follows then that squaring this to obtain an estimate of the power spectrum then results in a weighted quadratic combination of visibilities, where one multiplies the visibility from every baseline with that from every other baseline. These visibility product pairs are subsequently normalized to form their own individual estimates of the power spectrum before all the normalized pairs are summed together to form a final power spectrum. Measuring a power spectrum in this way combines the best aspects of the mapmaking and delay-spectrum approaches. Power spectra are formed directly by the cross-multiplication of visibility data, preserving the delay-spectrum approach's strategy of staying close to quantities measured by an interferometer; on the other hand, by multiplying together the data from every possible combination of baselines, one retains all the information from partially-coherent visibility pairs captured by mapmaking. Indeed, it can be shown \citep{liu_et_al2014a} that in the limit of infinitely fine Fourier space bins, the hybrid method is mathematically equivalent to the mapmaking approach. In practice, the hybrid method comes with the added benefit of allowing particular baseline pairs that are suspected of being affected by instrumental systematics to be more surgically downweighted. 
However, similar to map-making techniques, these approaches need much additional precision in calibration and modeling development in order to be implemented with confidence.

Fundamentally, the hybrid approach and the optimal mapmaking approach \citep{dillon_et_al2015a} are very similar. Whether it makes computational sense to examine all pairs of partially-coherent visibilities or to perform mapmaking as an intermediate data-compression step before power spectrum estimation (despite the complexity of the resultant map statistics) is an open question that depends on a number of factors, including the number of elements in the array, its layout, and its integration time.

\subsection{Calibration}
\label{sec:calibration}
As examined above, maximizing sensitivity to a limited number of spatial modes for the delay-spectrum technique requires a grid of antennas that simultaneously measures the same baselines using many redundant pairs of antennas. This redundancy may also be exploited to calibrate the array as well, as long as the true antenna positions fall close enough to that ideal grid \citep{liu_et_al2010}. This technique was pioneered with the MITEoR experiment \citep{zheng_et_al2014}, and has resulted in a package called OMNICAL developed for PAPER and HERA ( {https://github.com/jeffzhen/omnical}).  \citet{ali_et_al2015} provide a description of this technique as it applies to calibrating the PAPER array, which is mostly situated on a grid with many redundant baselines.  This section provides a high-level overview of such calibration as it applies to HERA.

Fundamentally, the problem of calibration is the determination of a complex, frequency-dependent gain for each antenna which arises due to differences in amplifiers, cables, etc. Given those gains, the visibility measured by antennas $i$ and $j$ at frequency $\nu$ is given by 
\begin{equation}
V_{ij}^\text{measured}(\nu) = g_i(\nu) g_j^*(\nu) V_{ij}^\text{true}(\nu), 
\end{equation}
where each $V_{ij}(\nu)$ is the measured or true visibility and $g_i(\nu)$ is the gain on antenna $i$ at frequency $\nu$.\footnote{This discussion ignores electromagnetic cross-talk between antennas and antenna-to-antenna variation of the primary beam.} With its highly-redundant array configuration (see \S\ref{sec:arrayConfig}), HERA's 350 antennas measure 61,075 visibilities, but only 6,140 unique baselines. Given a set of measured visibilities, it is possible to estimate both the 6,140 unique true visibilities and the 350 complex gains simultaneously because that system of equations is greatly overdetermined. A method for linearizing the system and applying standard minimum-variance linear estimators was developed by \citet{liu_et_al2010} and first applied in \citet{zheng_et_al2014}. Unlike traditional calibration of radio interferometers, redundant calibration makes no reference to any sky model, requiring merely that the sky is bright relative to the noise on each antenna (which is always true in the sky-noise dominated regime characteristic of 21\,cm observatories with wide fields of view). 

However, at each frequency, degeneracies in the redundant calibration procedure prevent it from solving four numbers per frequency for the whole array (down from 700 in HERA's case if full calibration, as opposed to redundant calibration, were needed). The first is an overall gain---for example, one can multiply each $g_i(\nu)$ by 2 and divide each $V_{ij}^\text{true}$ by 4 and get exactly the same measured visibilities. The others are an overall phase and two other phase terms that describe the tip and tilt of the array. These degeneracies mean that the output of any redundant baseline calibration also need to undergo a final, ``absolute'' calibration. 

\section{High-Level Requirements} 
\label{sec:requirements}
From the science and techniques discussed above, we may derive some high-level requirements predicated on optimizing for the delay-spectrum approach of detecting the EOR power spectrum while not unduly limiting other approaches or other science goals.  The fact that HERA is an experiment and not meant to be a long-term general use facility greatly facilitates this optimization. This section describes the element, configuration, frequency and sensitivity requirements that largely set the overall design concept.

\subsection{General Antenna Design and Configuration}
The primary new feature of HERA over previous generation experiments (in this context primarily PAPER) is the use of large elements to increase the sensitivity, as indicated in Figure~\ref{fig:boost}.  The cost-performance optimization will 
be discussed below, however we see that to address the central challenges of the delay-spectrum approach in the context of the wedge
HERA should use close-packed antennas that
minimize signal reflections over long delays and deliver significant forward gain relative to their horizon response.
Tests with prototype HERA antennas (Figs.  \ref{fig:delayspec} and \ref{fig:orbcommexptandbeammap}, discussed in \S\ref{sec:antenna})
indicate that a moderately large parabolic dish with a short focal height can meet these requirements
\citep{2016MNRAS.460.4320E,neben_et_al2016,patra_et_al2016,2016ApJ...825....9T}.  Note that to pursue map-making and hybrid techniques, additional antenna outriggers are valuable additions, so the design
should be able to accommodate $\approx$1~km baselines.  Note that in referring to the EOR sensitivity of the array, often the number of antennas just in the core are
indicated -- that is for EOR HERA-320 and HERA-350 are often used interchangeably.

\subsection{Frequency and Bandwidth}
\label{sec:freqs}
Figure~\ref{fig:cosmos} indicates the frequency range requirement to probe the expected timescale of the epoch of reionization using the 21cm line of hydrogen as the probe.  
These limits are derived from to-date complementary probes of reionization which include measurements of
the optical depth to last scattering in the CMB, QSO spectra, Ly-$\alpha$
absorption in the spectra of quasars and gamma ray bursts and the demographics
of Ly-$\alpha$ emitting galaxies
(Fig. \ref{fig:IonHist}). 
Constraints from these probes are still weak:
Ly-$\alpha$ absorption saturates at very small neutral fractions; galaxy
surveys directly constrain only the bright end of the luminosity function and
depend on an unknown escape fraction of ionizing photons to constrain
reionization; CMB measurements probe an integral quantity subject
to large degeneracies  Even when these observations are
combined into a single 95\% confidence region, the bounds remain weak.
For example, $x_{HI}$ spans almost the entire allowable range of [0,1]
at $z=8$. Twenty-one centimeter reionization experiments place much tighter constraints on ionization, with the
red band showing the forecasted 95\% confidence region derived from HERA data,
after marginalizing over astrophysical and cosmological parameters.

From these measurements and models, HERA is required to cover a redshift range of $z=6$ to 13, corresponding to a frequency range of 101--203 MHz.  We adopt 100--200 MHz for full performance as the requirement.  Extending the science to include the Dark Ages remains a goal so that efforts are on-going to design a feed to increase the lower limit down to 50 MHz without compromising the performance in the above requirement range.  The fall-back position for low frequencies is a serial deployment of a scaled version of the HERA feed or to potentially build additional elements specifically for low frequencies.  

The scientific requirement on channel bandwidth is to allow access to k-modes of $\sim 1\,h$~Mpc$^{-1}$, which requires about 256 channels over the 100 MHz bandwidth.  However, in order to handle radio frequency interference as well as to allow the bandpass to be characterized, a specification of 1024 channels has been chosen.  This yields a channel bandwidth of 97.7 kHz.

The total simultaneously processed bandwidth is the full 100 MHz, since we wish to efficiently probe the entire redshift range.  The main impact from this requirement is the bandwidth from the digitizer back to the correlator, however this bandwidth is easily handled with current generation digital back-ends and will be greater than 100 MHz.

\subsection{Delay Response}
\label{sec:delayspec}
Early analysis \citep{elementmemo} indicated that to trace the EOR over a wide range of redshift, internal reflections should be attenuated by about 60 dB from the initial incoming wave by about 60 ns, or a voltage standing wave ratio (VSWR) $<$ 1.002 for frequencies $<$ 17 MHz.  This was a conservative value based on estimates of power in spectrally smooth foreground sources \citep{santos_et_al2005,2008MNRAS.385.2166A, deoliveira2008, jelic_et_al2008, bernardi_et_al2010} and experience with PAPER dealing with foreground systematics.
This provided the basis for the initial design of the element.    For the contemplated feeds on a primary focus antenna, it was found that a focal length/diameter radio ($f/D$) of approximately 0.32 was optimal.  We may therefore estimate the attenuation as a function of delay and diameter and assign this a ``cost function''.  

\begin{figure}
\centerline{
\includegraphics[width=0.48\textwidth]{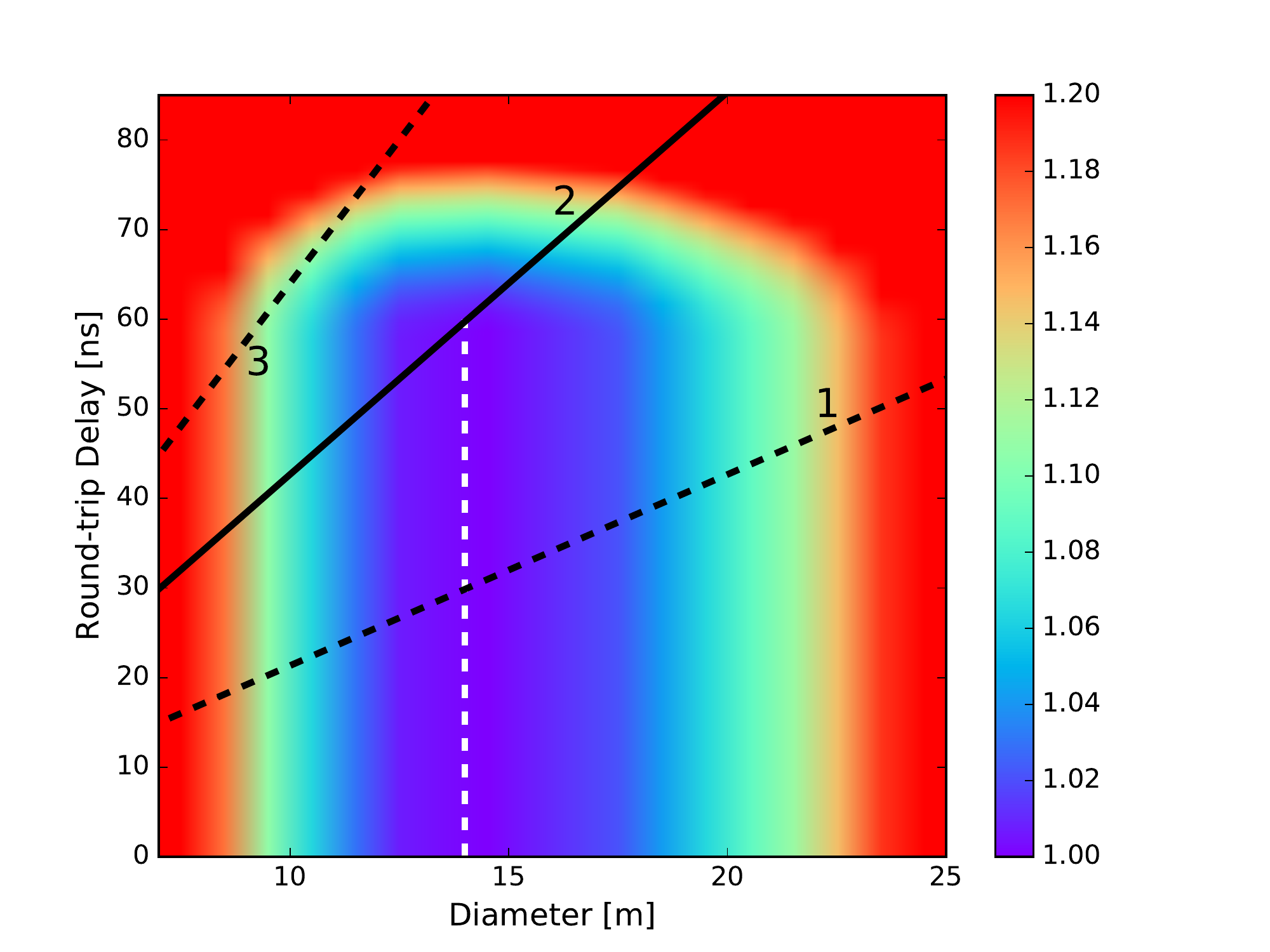}
}
\caption{\small The background coloring with colorbar shows the relative cost/performance for delay and cost at a fixed performance as a function of element diameter.  See \S\ref{sec:delayspec}, \S\ref{sec:cost} and \cite{elementmemo} for details.  A diameter of 14-m (vertical dashed line) is near the cost minimum and consistent with assumptions for the delay specification.
\label{fig:costfig}}
\end{figure}

Figure \ref{fig:costfig} summarizes the adopted cost function augmented with an actual cost scaling.  
The vertical axis shows a steep delay ``cost'' at 60 ns and the horizontal axis shows the cost/performance model as a function of element diameter, as described in section \ref{sec:cost}..
The color coding is the product of the two assigned costs, with red showing a high cost normalized to 350 14~m elements (clipped at a ratio of 1.2) and purple being the minimum.
The black lines labelled `1', `2' and `3' are the round-trip travel times for 1, 2 and 3 reflections between the feed and the vertex.

To try and obtain 60 dB by 60 ns we may interpret the black delay lines in Fig. \ref{fig:costfig} in one of two ways:  
(a) a fixed attenuation/reflection determines what diameter is required, and
(b) a fixed diameter determines how much attenuation per reflection can be allowed.
Interpreting Fig. \ref{fig:costfig}, for one feed-vertex-feed round-trip and the assumed $f/D=0.32$ over the diameters of interest, all delays are less than 60 ns.  We therefore don't have to reduce the reflections to the extreme amount needed to get to 60 dB of attenuation within one round-trip.   
If we want to be able to handle two round-trips worth of delay, the diameter should be less than 14.2~m, which is near the cost minimum.  Alternatively, if we assume 14.2~m, then we can estimate that each reflection must average a return loss better than 15 dB, an aggressive target.  And finally, if we need to allow for three round-trips, the antenna diameter would have to be less than 9.4~m, where the cost is quickly rising.  To handle the delay requirement therefore, the diameter should be less than 14.2~m, or the return loss specification becomes very aggressive.

As a key specification, this has been a focus of study with a series of prototypes and analyses based on modeling, and measurements have been conducted to provide much more rigor than this early specification \citep{2016MNRAS.460.4320E,neben_et_al2016,2016ApJ...825....9T,patra_et_al2016}.  The simple specification set above more appropriately becomes a series of curves for attenuation and delay at different wavenumbers for realistic foreground model, as discussed below.

\begin{figure}[h!]
	\centering
    \includegraphics[width=0.49\textwidth]{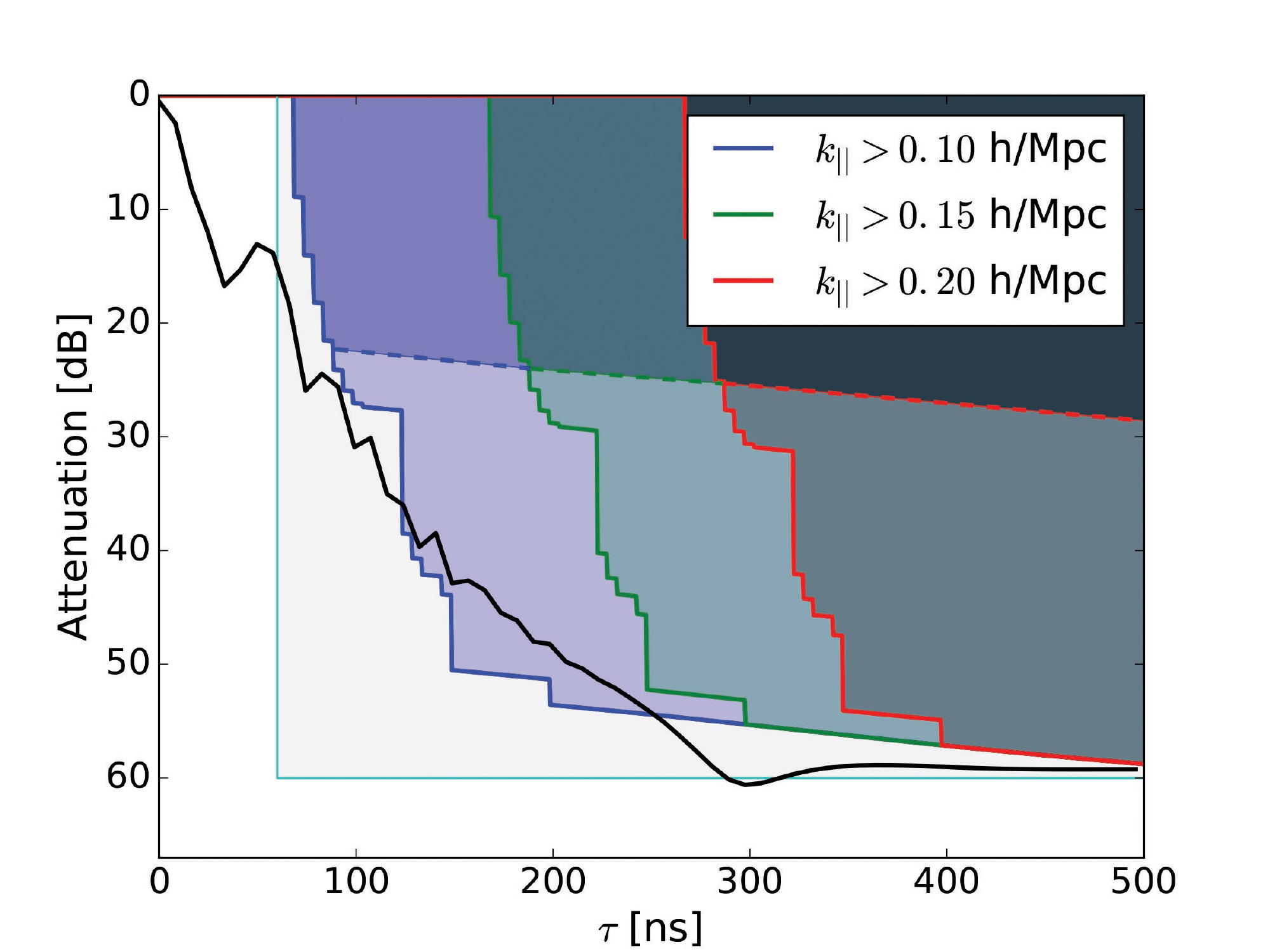} 
	\caption{Antenna delay attenuation specification with realistic foregrounds from \cite{2016ApJ...825....9T}.  
	The blue, green and red lines show the maximum attenuation needed at a given delay for the foregrounds to be below the expected EOR signal level for different $\kpar$.  The lower lines are for the worst case with no additional processing, while the dashed line assumes inverse covariance weighting.
	The teal line is the 60 dB by 60 ns initial conservative specification.
	 HFSS calculations (black line) for the expected attenuation show the HERA dish meeting the more sophisticated specification for $\kpar\gtrsim0.15$ h~Mpc$^{-1}$.
}
	\label{fig:delayspec}
	\vspace{-10pt}
\end{figure}

The more detailed specification is shown in Fig. \ref{fig:delayspec}, adapted from \cite{2016ApJ...825....9T}, showing the modeled results at $k$ = 0.10, 0.15, and 0.20 $h$~Mpc$^{-1}$ (blue, green and red respectively) for the adopted design for two different cases.  If the system attenuation as a function of delay (black line in figure) is greater than the attenuation needed to measure a specific $\kpar$ based on a realistic foreground model, the measurement should be free from foregrounds at the level of the expected EOR power for those wavenumbers and larger.  The case extending down to $\sim 55$ dB is  the worst case for the raw system with no additional processing.  The dashed line at $\sim 25$ dB is for application of an inverse covariance weighted optimal quadratic estimator  of the data \citep{patra_et_al2016}.
The analysis shows that for the worst case
the dish design should allow for EOR detections for $k\gtrsim0.15$ h~Mpc$^{-1}$, while covariance weighting should allow detections at  $k\gtrsim0.10$ h~Mpc$^{-1}$.  We also see that the initial simple 60 dB by 60 ns specification was overly conservative (shown as the lightly shaded area bounded by the teal line), however it served as a good driving goal.

\subsection{Sensitivity Optimization}
\label{sec:cost}
As an instrument to characterize the power spectrum over the evolution of the EOR, the specification on sensitivity is to make at least a nominal detection over the redshift region of support ($z=6-13$) along with a very robust detection at the peak.  As seen in Fig. \ref{fig:Sensitivities} and Table \ref{tab:signif}, an array comprising 320 14-m core antennas (labelled HERA-350, which includes the outriggers) has sensitivity to a
fiducial reionization model across that redshift range.  We therefore adopt the delay-spectrum power spectrum sensitivity of 320 14-m close-packed antennas as the minimum sensitivity.  
For this sensitivity, the number of elements needed for a given diameter is
\begin{equation}
\label{eq:ND}
N_{\text core} = (320*14)/D
\end{equation}
where $N_{\text core}$ and $D$ are the number of antennas in the core and their diameter respectively.  Note that the linear number-diameter dependence was derived by running multiple sensitivity codes ({\em e.g.}  {21cmSense}) over a range of values for the HERA configuration for wavenumbers greater than 0.15~$h$~Mpc$^{-1}$.  This dependency differs from \cite{2013ExA....36..235M} and analytical expressions, mostly likely due to the assumed configuration.
Optimization is therefore done for this sensitivity against cost for various diameters and total element numbers, while trying to minimize multiple reflections on long delays.   

To determine this optimal diameter, full system costings were done on a range of diameter sizes from 5~m - 25~m, with the total number set by Eq. \ref{eq:ND}.  This included the full system, except the post-processing (that is, everything after the archived correlator output). 
The costing information in Fig. \ref{fig:costfig} is shown as the colored background oriented along the horizontal axis, recalling that the total ``cost'' function is the product of the actual costing and the delay-attenuation curve.
The resulting normalized system cost/performance-curve  becomes very flat (colored dark blue/purple) for diameters between about 12 and 15 meters, which is consistent with the chosen value of 14 meters based on other system considerations.

\section{System Design} \label{sec:design}
\begin{figure*}
	\centering
	\includegraphics[width=.95\textwidth]{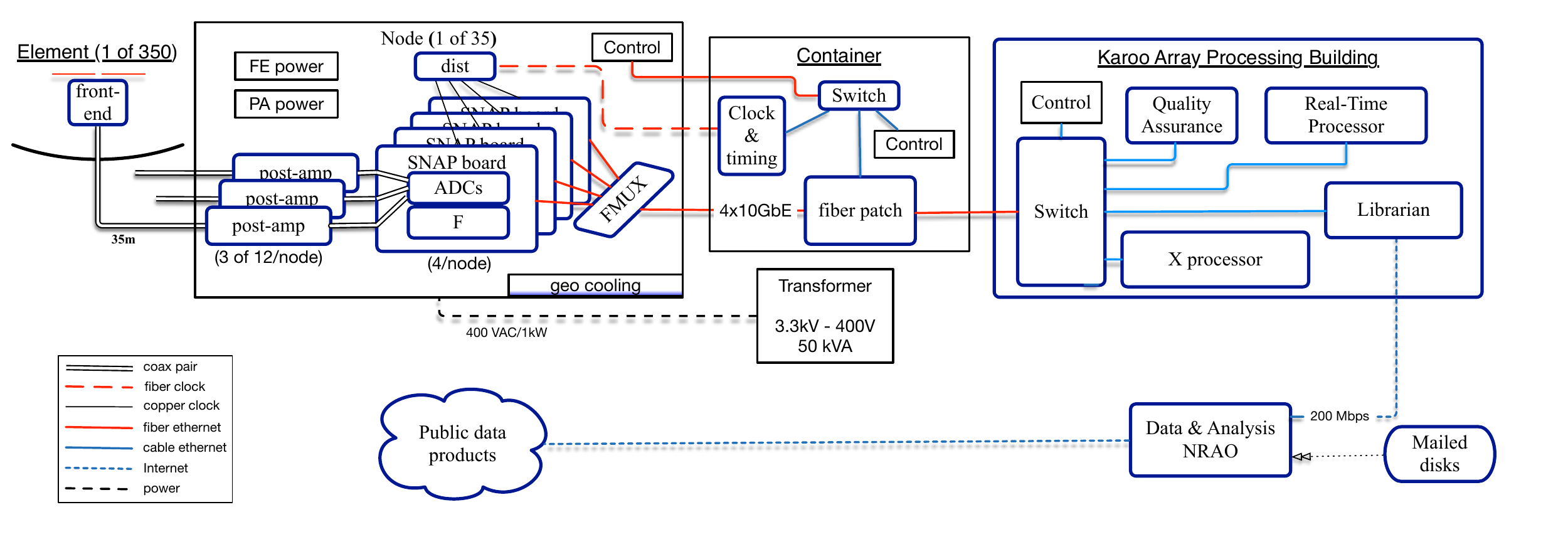}
	\caption{HERA's signal path.  Front-end amplifiers at the antenna feed drive signals on short coaxial cables to 
field nodes.  Nodes contain post-amplifiers and Smart Network ADC Processor (SNAP) boards that digitize, channelize,
and packetize data for optical transmission in 10 Gb Ethernet format.  Optical fibers are aggregated in a field container
onto a 10 km fiber bundle connecting to the Karoo Array Processing Building, where signals are cross-multiplied
in the X processor.  After correlation, visibilities are stored by the Librarian, pre-processed and redundantly calibrated
by the Real-Time Processor, and transmitted over the network to clusters for storage and analysis.  Final products are
hosted on public-facing NRAO servers, with a web interface for selecting and downloading data.}
	\label{fig:overallBlockDiagram}
\end{figure*}

As described in the previous section, the critical insights from the first generation 21\,cm EOR experiments have been applied to define
the requirements for HERA---an instrument designed to ensure that foregrounds remain bounded
within the wedge while delivering the sensitivity for
high-significance detections of the 21\,cm reionization power spectrum with
established foreground filtering techniques
\citep{pober_et_al2014,greig_mesinger2015}.
In this section, we summarize key features of the HERA design (see Table~\ref{tab:BasicParameters}) 
and system architecture (see Fig.~\ref{fig:overallBlockDiagram}).
This architecture directly inherits from the PAPER and MWA experiments.  HERA begins by reusing the
analog, digital, and real-time processing systems deployed for PAPER-128.  
This allows for immediate observing with the new elements with a well-characterized system.
As HERA develops, this
architecture is incrementally upgraded to improve performance and add features while simultaneously
addressing issues of modularity and scalability.  As with PAPER, HERA proceeds in stages of development,
with annual observing campaigns driving a cycle of development, testing, system integration, calibration, and analysis.
This cycle ensures that HERA's instrument is always growing, that systematics are being found and eliminated at the
earliest build-out stages, that data analysis pipelines are tested and debugged while data volumes are smaller,
and that HERA is always producing high quality science.

Note that in addition to the 19 elements currently observing in South Africa (and an initial prototype near Berkeley, CA), a pair of elements at the Green Bank Observatory in West Virginia have been used extensively in validation testing.  Also, three elements are currently under construction at the Mullard Radio Astronomy Observatory outside Cambridge, UK for additional feed and element tests.  This allows for an independent testing platform as the group at Cambridge continues to investigate both a broad-band feed, as well as ways to improve the current analog system for an improved match and noise performance.

For details of the first generation HERA system (the existing and commissioned PAPER signal path), see \cite{parsons_et_al2010}.  This section will discuss the new system to be deployed beginning and 2016 and tested in conjunction with the previous system.  Figure~\ref{fig:overallBlockDiagram} provides a system block diagram of the new HERA architecture.

\begin{table*}
\small
\Caption{0.0in}{0.99}{-0.1in}{HERA-350 design parameters and their observational consequences. Angular scales computed at 150\,MHz.}
\label{tab:BasicParameters}
\begin{center}
\begin{tabular}{l | l}
\multicolumn{1}{c}{\emph{\textbf{Instrument Design Specification}}} & \multicolumn{1}{c}{\emph{\textbf{Observational Performance}}}\\
\hline
\textbf{Element Diameter:} 14\,m & \textbf{Field of View:} 9\arcdeg \\
\textbf{Minimium Baseline:} 14.6\,m & \textbf{Largest Scale:} 7.8\arcdeg\\
\textbf{Maximum Core Baseline:} 292\,m & \textbf{Core Synthesized Beam:} 25\arcmin\\
\textbf{Maximum Outrigger Baseline:} 876\,m & \textbf{Outrigger Synthesized Beam:} 11\arcmin\\
\textbf{EOR Frequency Band:} 100--200\,MHz & \textbf{Redshift Range:  } $6.1 < z < 13.2$ \\
\textbf{Extended Frequency Range:} 50--250\,MHz & \textbf{Redshift Range:} $4.7 < z < 27.4$ \\
\textbf{Frequency Resolution:} 97.8\,kHz & \textbf{LoS Comoving Resolution:} $1.7$\,Mpc (at $z=8.5$)\\
\textbf{Survey Area:} $\sim 1440$ deg$^2$ & \textbf{Comoving Survey Volume:} $\sim 150$\,Gpc$^3$ \\
$\mathbf{T_\textbf{sys}}$: $100 + 120 (\nu/\rm{150~MHz})^{-2.55}$ K & \textbf{Sensitivity after 100\,hrs:} 50 $\mu \rm{Jy}~\rm{beam}^{-1}$  \\
\hline
\end{tabular}
\end{center}
\end{table*}

\subsection{Site}
\label{sec:site}
HERA is located at the Karoo Radio Astronomy Reserve, one of the two selected sites to host the SKA telescope. The Karoo Radio Astronomy Reserve was established in 2007 through the Astronomy Geographic Advantage Act in order to provide the preservation of its radio--quite environment by restricting the use of certain radio frequencies and limiting the transmitting power in an area of $\approx 160$~km radius centred at the SKA core site, near to the town of Carnavon, in the Northen Cape province. The level of protection is set to meet the requirements of the SKA project\footnote{http://skatelescope.org}. Besides offering an exceptionally good RFI environment, the Karoo site is still very accessible, making it an ideal hosting site for radio instrumentation. Beside the already mentioned HERA and SKA, it hosts the 7--element Karoo Array Telescope (KAT--7; \citealt{2016MNRAS.456.1259B}), the 64--antenna MeerKAT array (currently under construction; \citealt{2012AfrSk..16..101B}) and the CBASS telescope \citep{2010AAS...21538702S}. Since HERA is a scientific pathfinder for the SKA on an SKA site, it has been designated an SKA Precursor instrument.

A key element of site infrastructure is the Karoo Array Processor Building (KAPB) that hosts the storage and computing for the radio instrumentation deployed on site. In particular, the KAPB hosts the correlator for the KAT--7 and MeerKAT arrays. After correlation, the data are also archived at the KAPB and later streamed off site via fibre connection to the Cape Town office. As discussed in Section~\ref{sec:digital}, the KAPB also hosts the correlator and archive for HERA.

\subsection{Antenna}
\label{sec:antenna}
The goals of the design principles are three-fold:
(1)optimize for the delay-spectrum technique of measuring the EOR power spectrum,
(2) minimize costs, and
(3) design for a limited lifetime of about five years.
The first item primarily means that chromatic effects corresponding to delays appropriate for the measurement described above must be below the expected signal level, which essentially determines the focal length.  The second item constrains the diameter and element count, as well as the focal length over diameter ratio ($f/D$), based on a cost function and maximizing sensitivity per element.  And the third item constrains the construction materials and methods and the operational model.  This led to a fixed transit element with a diameter of 14 meters to strike an optimal balance between sensitivity and systematics, as discussed previously.

The large collecting area of the HERA element yields nearly 5 times the sensitivity of an MWA tile and more
than 20 times that of a PAPER element, but it does so without substantially
degrading the ability to isolate and remove foreground emission on the basis of
spectral smoothness.  As shown in
\Mycitet{parsons_et_al2012b} and discussed earlier, the amplitude and timescale of signal reflections
relates directly to the leakage of smooth-spectrum foregrounds into regions of Fourier space 
used to measure reionization, as discussed in \S\ref{sec:wedge}.
To facilitate foreground filtering, HERA's antenna element is designed to suppress reflections at long time delays.

\begin{figure*}
\centering
\includegraphics[width=.4\textwidth]{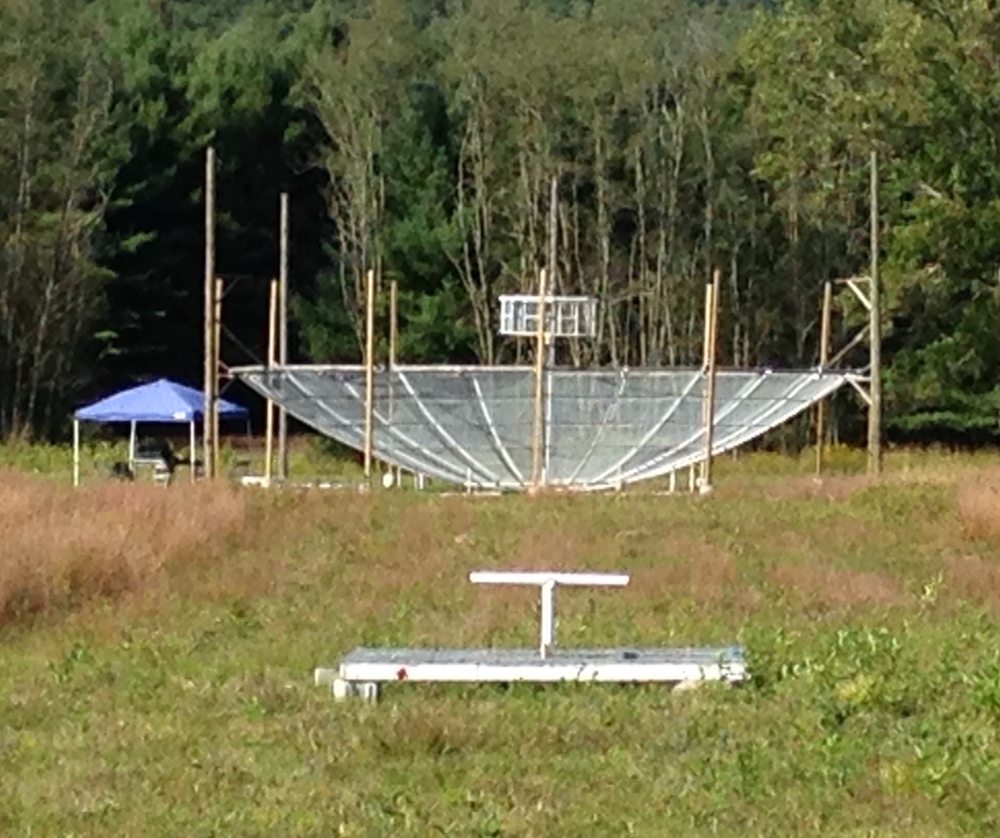}
\includegraphics[width=.4\textwidth]{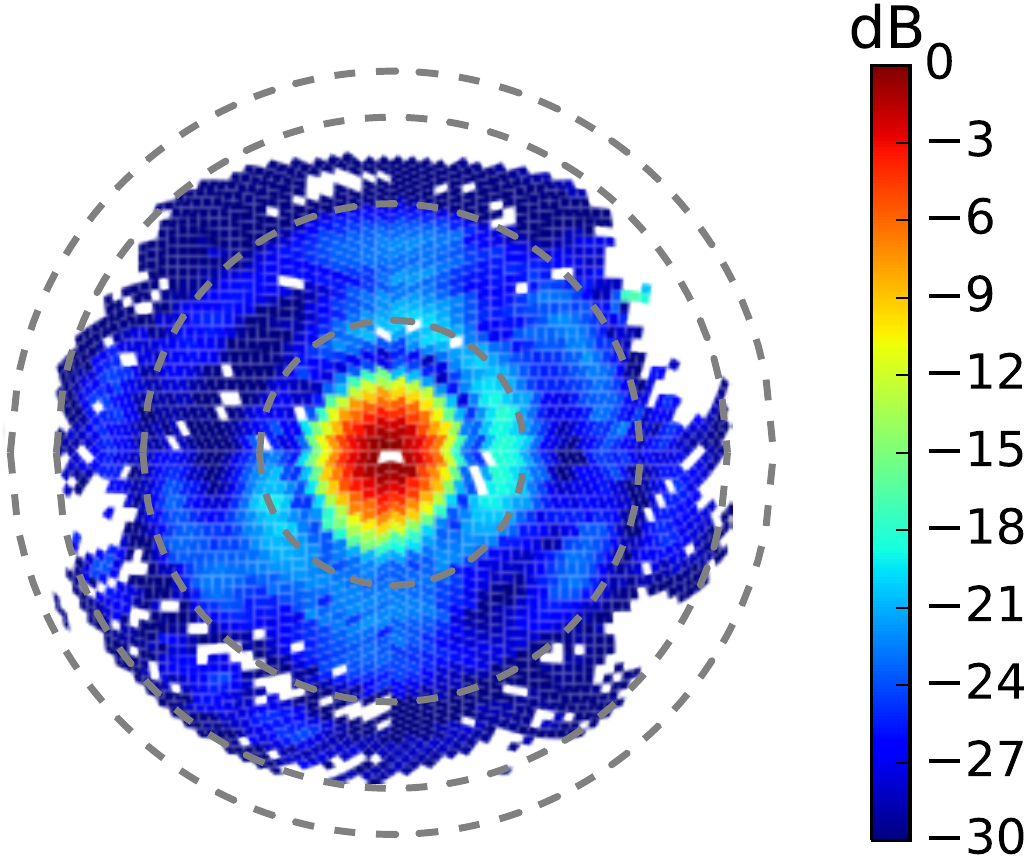}
	\caption{Left: The first of two prototype dishes at NRAO--Green Bank, used for measuring beam frequency structure with reflectometry and the beam pattern at 137\,MHz by comparing satellite signals to the reference dipole in the foreground \citep{neben_et_al2016}. Right: The measured EW power pattern plotted with dashed lines marking zenith angles of 20$^\circ$, 40$^\circ$, 60$^\circ$, 80$^\circ$.} 
	\label{fig:orbcommexptandbeammap}
\end{figure*}

Figure~\ref{fig:orbcommexptandbeammap} shows the beam pattern measured at 137 MHz with a beam mapping
system using the ORBCOMM satellite network \Mycitep{neben_et_al2016}.  Results
indicate an effective per-element collecting area of 93\,m$^2$, or about 60\% efficiency for this version.  The measured primary beam is
consistent with simulations at the 0.1\% to 0.5\% level, with a full width at
half maximum of $\sim$$10^\circ$, and a first sidelobe at -20 dB \citep{ewallwice_et_al2016b, neben_et_al2016, patra_et_al2016,2016ApJ...825....9T}.
Feed-to-feed coupling between the two adjacent antennas in Green Bank has been measured to be below -50 dB, providing confidence that mutual coupling will be  
intrinsically and algorithmically manageable.  Studies of elements in a bigger array and potential screening approaches are also underway.

\subsubsection{Element Construction}
To minimize cost and aid in construction in a remote location, the elements are built from readily available standard construction materials, making for a very cost-effective design.  The feed is supported from three utility poles using line and hardware primarily from boating activities which can handle the load and external environment.  Given the close-packed design, each pole (except for the perimeter) is shared between three antennas.  This both reduces the number of poles, but also makes for a balanced load.  Note that the perimeter poles can be stayed with guy lines if needed.   The dish rim is also supported off of these poles, each with three smaller posts in between such that the outer perimeter is regular dodecagon shape.  Six of the nine intermediate posts are shared with adjacent antennas.

The center of the antenna is a cast-in-place concrete donut-shaped hub with 12 sleeves for PVC spars which provide the support for the surface and 12 horizontal spars which allow for additional support.  The twelve surface spars are made of 60~mm PVC pipe, which are stressed along three support points to approximate a parabolic shape.  Note that an ideal moment-loaded beam actually attains a true parabola.  The three support points (one at the hub, one about 3 meters out radially and one at the rim-end) are at the correct height and angle for the underlying parabola and the PVC essentially acts as a spatial filter.

\begin{figure}
\centerline{
\includegraphics[height=4cm]{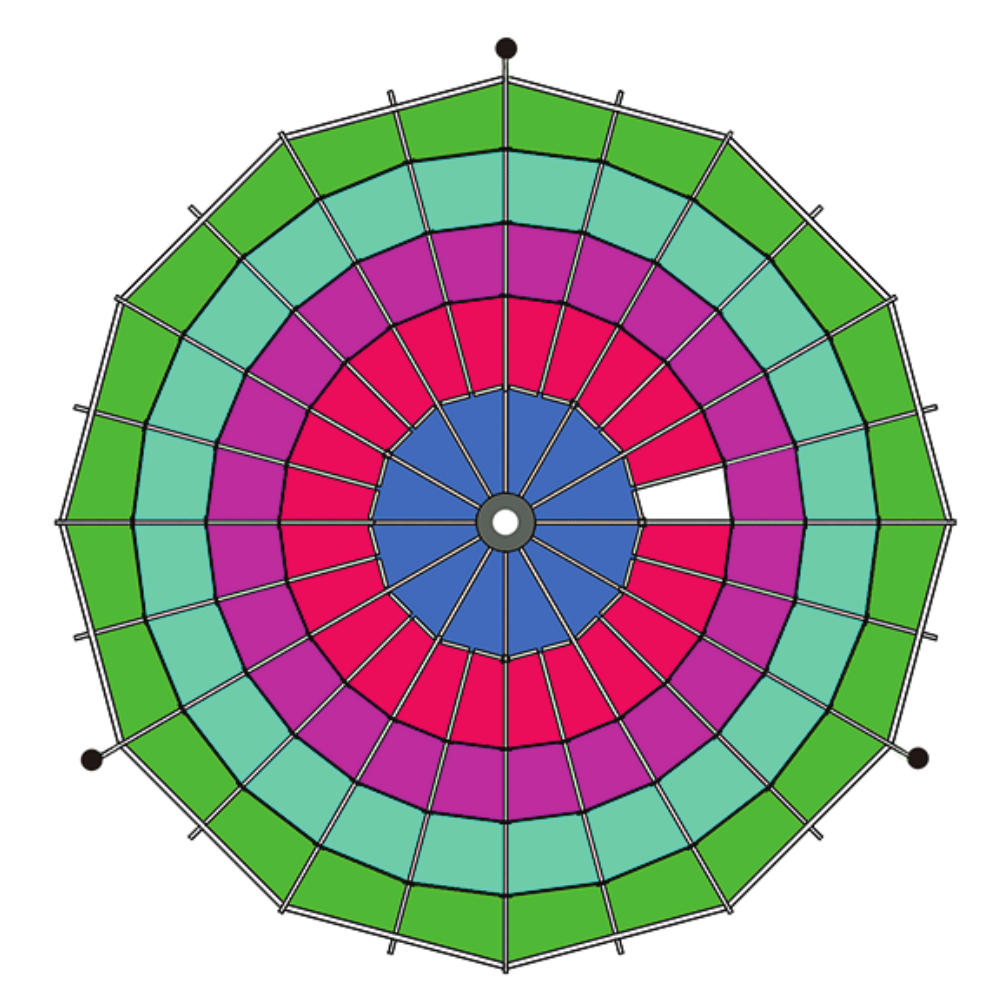}
\includegraphics[height=4cm]{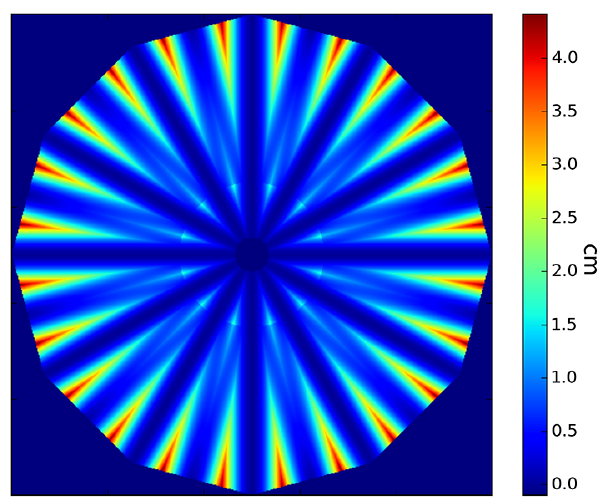} 
}
\caption{\small Element panel plan (left) and residual from a true paraboloid (right). The white panel denotes the location of the door.
\label{fig:elementplan}}
\end{figure}
The shape between the spars is actually a parabola rather than a paraboloid, so the dish surface may more properly be called a ``faceted parabola''.  If only twelve spars were used for the entire surface, simple Ruze losses would be quite high (about 10\% at 150 MHz).  Therefore, at a radius of 2.4 meters another horizontal member is placed, which launches another parabolic spar leading to a Ruze loss less than 1\%. Fig.~\ref{fig:elementplan} shows this surface scheme and the panelization (left) and the offsets (in mm) of that scheme from an ideal parabola.  The panel that is ``left out'' is the location of a door to allow access into the hub (via a small removable door and bridge).  The additional intermediate spars also make the panel sizes much more manageable and better secured.  The panel arrangement is also set by a standard width of mesh roll of 1.22~m.

\subsubsection{Feed}
\label{sec:feed}
The feed design is proceeding in three phases:
\begin{itemize}[leftmargin=0.7in]
\item[\em Phase 1:]  use existing PAPER dipole with new backplane and existing PAPER signal path,
\item[\em Phase 2:]  updated backplane with new matching network and analog signal path,
\item[\em Phase 3:]  broad-band or separate low-frequency feed for full spectral coverage.
\end{itemize}
These will be discussed in turn.

\begin{figure*}
\centerline{
\includegraphics[width=0.95\textwidth]{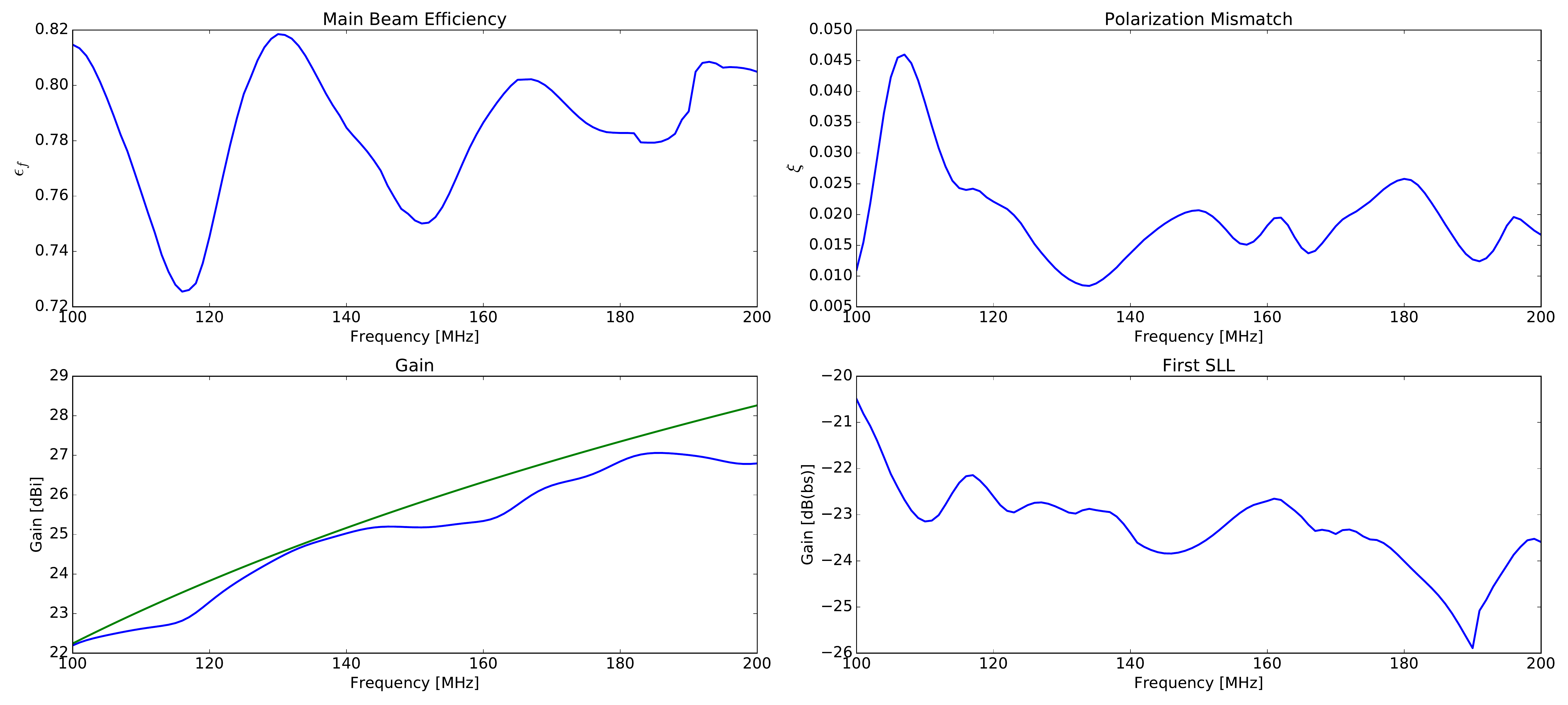}
}
\caption{\small Antenna parameters from the feed optimization analysis.  All calculations were done using HFSS on the final design as described.  {\bf Upper left:}  fraction of feed power illuminating the primary (main beam efficiency).  {\bf Upper right:}  polarization mismatch as defined in Eq.  \ref{eq:polarmismatch}.  {\bf Lower left:}  total gain (blue) contrasted with a fixed efficiency of 78\%.  {\bf Lower right:}  first side-lobe level relative to the main beam peak.}
\label{fig:ant_params}
\end{figure*}

\noindent
{\bf Phase 1:}  Initially, the feed reuses the original PAPER sleeved dipole, however in a new optimized cylindrical backplane configuration.  As discussed in \cite{feedmemo}, the optimization was based on (a) main beam efficiency, (b) cross-coupling (integrated feed pattern on the dish), (c) standing waves (blockage size), (d) frequency response and (e) polarization match.  The design parameters were size of backplane, height of mast, and height of cylinder.  A range of conical structures were looked at but were ruled out since it was found they introduce additional frequency structure with no real improvement in performance.  The polarization figure-of-merit shown in Figure~\ref{fig:ant_params} is a simplified variant of the fractional power leakage as defined in \cite{moore_et_al2016}:
\begin{equation}
\xi = \frac{\integral_{-\pi}^{\pi}\left[\sqrt{G(\theta,\phi=0)} - \sqrt{G(\theta,\phi=90)}\right]^2\sin\theta d\theta}
               {\integral_{-\pi}^{\pi}\left[\sqrt{G(\theta,\phi=0)} + \sqrt{G(\theta,\phi=90)}\right]^2\sin\theta d\theta}
     \label{eq:polarmismatch}
\end{equation}
where $G$ is the beam pattern.  This is basically a measurement of the beam match in the E and H planes of the antenna response.
The phase 1 design process and outcome is documented in  \cite{feedmemo} and is currently under test on the 19 elements in South Africa. Additional studies are on-going at Green Bank and Cambridge.

\begin{figure*}
\centerline{
\includegraphics[width=0.32\textwidth]{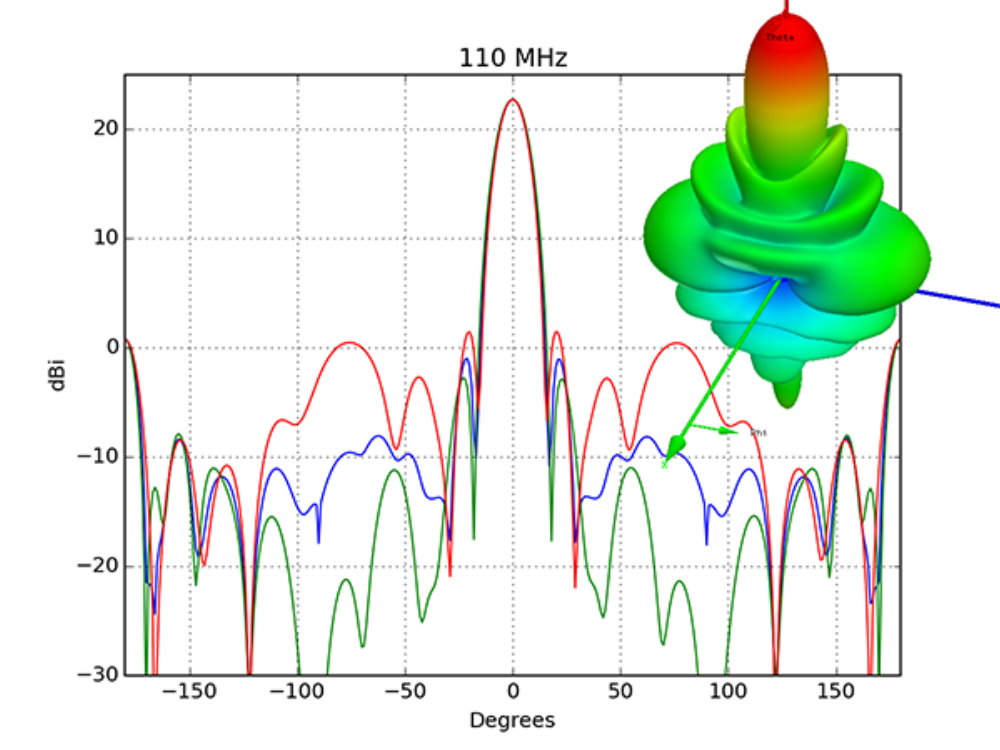}
\includegraphics[width=0.32\textwidth]{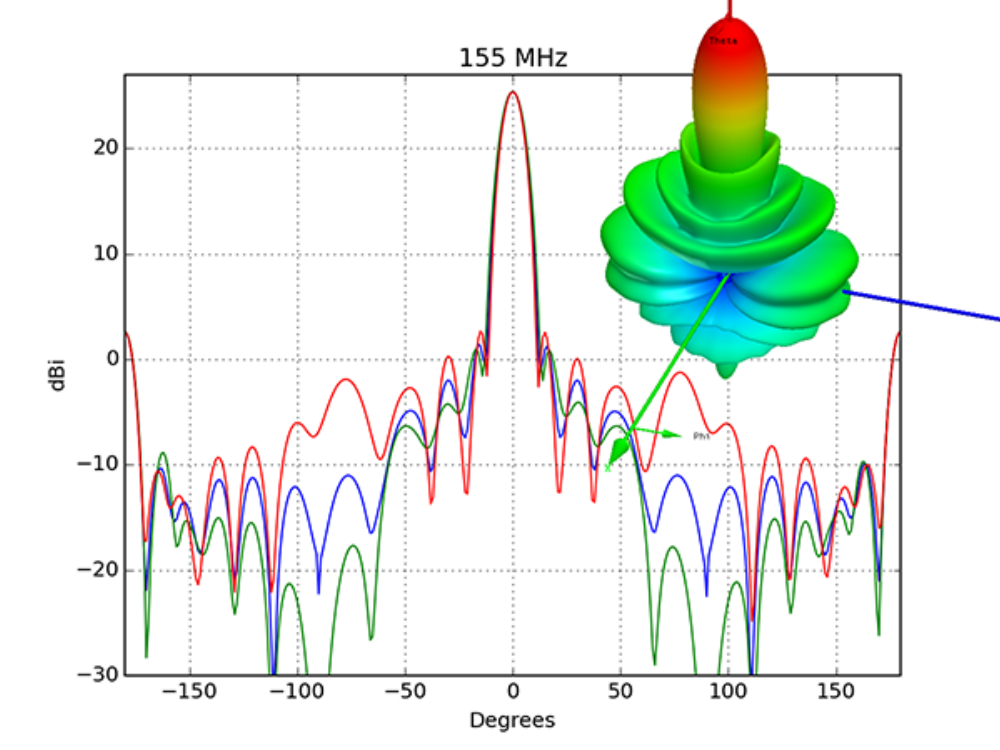}
\includegraphics[width=0.32\textwidth]{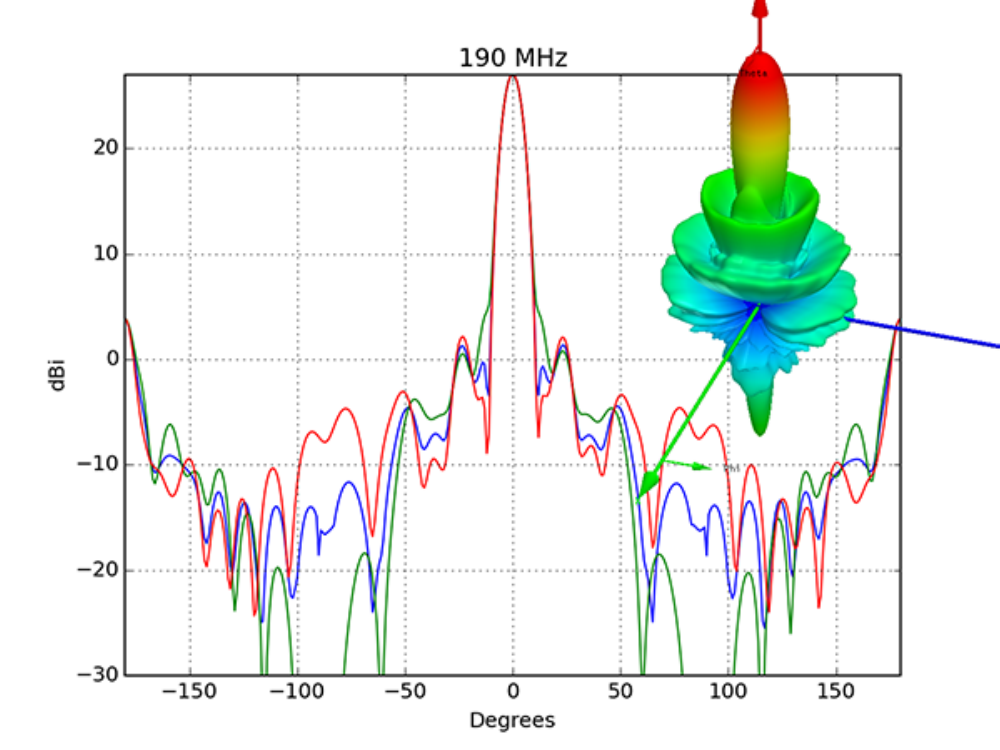}
}
\caption{\small Antenna patterns at 110 MHz, 155 MHz and 190 MHz calculated with HFSS and the adopted design.  Red curves are E-plane, green are H-plane at blue at 45$^\circ$.  Inset is the full 3D pattern.}
\label{fig:beampatterns}
\end{figure*}

This first phase design uses a 172~cm diameter backplane, with a 36~cm cylinder height.  The dipole is held at 36~cm from the backplane and the back of the feed is rigged at 4.9~m above the vertex of the dish.  The 10~dB input match for the feed itself spans from about 105 - 200 MHz.  The resultant HFSS-calculated antenna performance parameters are shown in Fig.~\ref{fig:ant_params} and beam-patterns for 110~MHz, 155~MHz and 190~MHz are shown in Fig.~\ref{fig:beampatterns}.  Note that this feed differs from the earlier ORBCOMM measurements, which used a pre-optimized version.

The feed is supported by spring-tensioned aramid fiber lines from the three poles via a rigid support to which the feed cage and dipole attach.  The tensioning line goes down centrally to the hub to hold the feed rigidly in place at the correct height.  This also stabilizes the feed in position and allows for strain relief on the coaxial cables.  

The low noise amplifier/balun integrates directly to the base of the feed dipole in the same configuration as per PAPER.  The 150~m length of 75 $\Omega$ cable used in the initial design also runs down centrally and then over to the post-amplifier module, which is housed near the RFI-shielded container containing the digital electronics.  

\noindent
{\bf Phase 2:}
The next phase replaces the extant PAPER signal path with new front-end and post-amplifier modules, discussed in \S\ref{sec:analog}.  Depending on the status of feed optimization and development the current feed backplane or the feed itself may be swapped out with new designs.  The new system incorporates the node-based system, with short analog cables and field-deployed digitizers (\S\ref{sec:digital}).

\begin{figure}
\centerline{
\includegraphics[height=0.35\textwidth]{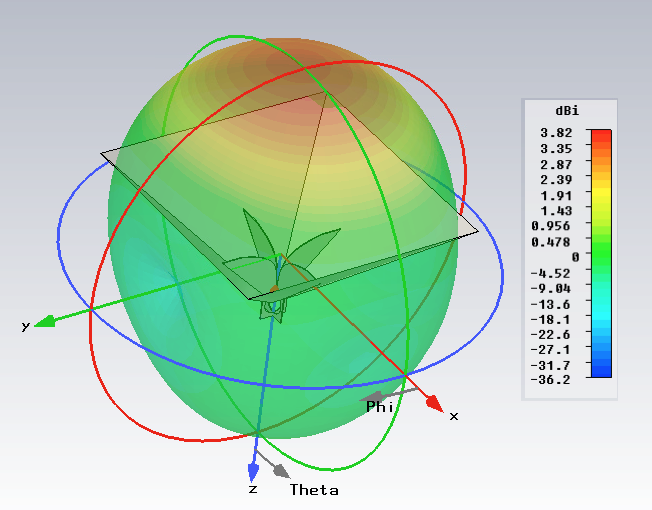}
}
\caption{\small Future potential HERA feed showing its directive beam pointed towards the dish. It is based on a modified ridge TEM horn and is currently under investigation at Cambridge University.}
\label{fig:futureFeed}
\end{figure}

\noindent
{\bf Phase 3:} 
A wideband feed is being designed for the next phase of HERA (see {\em e.g.} Fig. \ref{fig:futureFeed}). This feed will have wider bandwidth (50-250 MHz) and have comparable (or better) performance across the EOR band; if not the deployment will be after EOR observations. This wider band will open the window of Cosmic Dawn and the latest stages of the Epoch of Re-ionization to HERA. Observations at frequencies above 200 MHz can also provide a consistency check for lower frequency measurements, as the maximum brightness of the 21 cm signal at late times can be constrained with other probes \citep{pober_et_al2016b}. The feed version shown is based on a modified TEM horn concept where the feed's beam is optimized for the appropriate illumination of the HERA dish to maximize sensitivity. This is to maximize the effective aperture while minimizing the system temperature (receiver noise, ohmic losses, spill-over, etc.).  The new feed will replace the current PAPER dipoles and will therefore be mechanically compatible with the current dishes being deployed in South Africa.  

Furthermore, as shown in \cite{2015ExA....39..567D}, room temperature ultra low receiver noise ($<$35~K including matching noise) above ~100 MHz can be achieved with COTS transistors if a proper feeding mechanism and matching are designed. Despite the fact that sky noise is dominant for most of the EoR band, good matching is always critical in wide band systems, and specifically in the HERA case it is important to ensure good power match as well as noise match in order to reduce the effects of unwanted reflections \citep{fagnoni_delera_iceaa2016}.
We envisage having such a feeding mechanism in the new feed where the first stage LNA will be connected directly to the feeding point of the feed antenna. 

Additionally, we are analyzing the effect of coupling with numerical simulations where we can quantify and understand the effects of mutual coupling, which so far only looks substantial in the far out side-lobes. This will allow us to optimize the feed design for reduced coupling as well as evaluate the need of ``skirts'' for cross-coupling reduction or other further dish optimizations. This work will benefit from the existence of several prototype systems including the 3-dish HERA system near Cambridge.

\subsection{Array Configuration}
\label{sec:arrayConfig}

As a focused experiment, HERA's configuration is optimized for the robust foreground-avoidance approach to EoR power spectrum estimation. 
This requires a densely-packed core of antenna elements, maximizing sensitivity to short baselines---those least contaminated by foreground chromaticity ({\em i.e} the wedge). 
Furthermore, the configuration should be highly redundant, both to increase sensitivity for the delay-spectrum strategy of PAPER \citep{parsons_et_al2012a} and provide for redundant baseline calibration \citep{liu_et_al2010,zheng_et_al2014}.
On the other hand, it is also desirable to improve the mapmaking ability of the array for calibration and map-based power spectrum estimation with longer baselines and a densely-sampled $uv$-plane \citep{dillon_et_al2015a}. This tradeoff was closely examined by \citet{dillon_parsons2016} and the HERA configuration was drawn from one of the designs considered therein.

\begin{figure*}
	\centering
	\includegraphics[width=1\textwidth,clip]{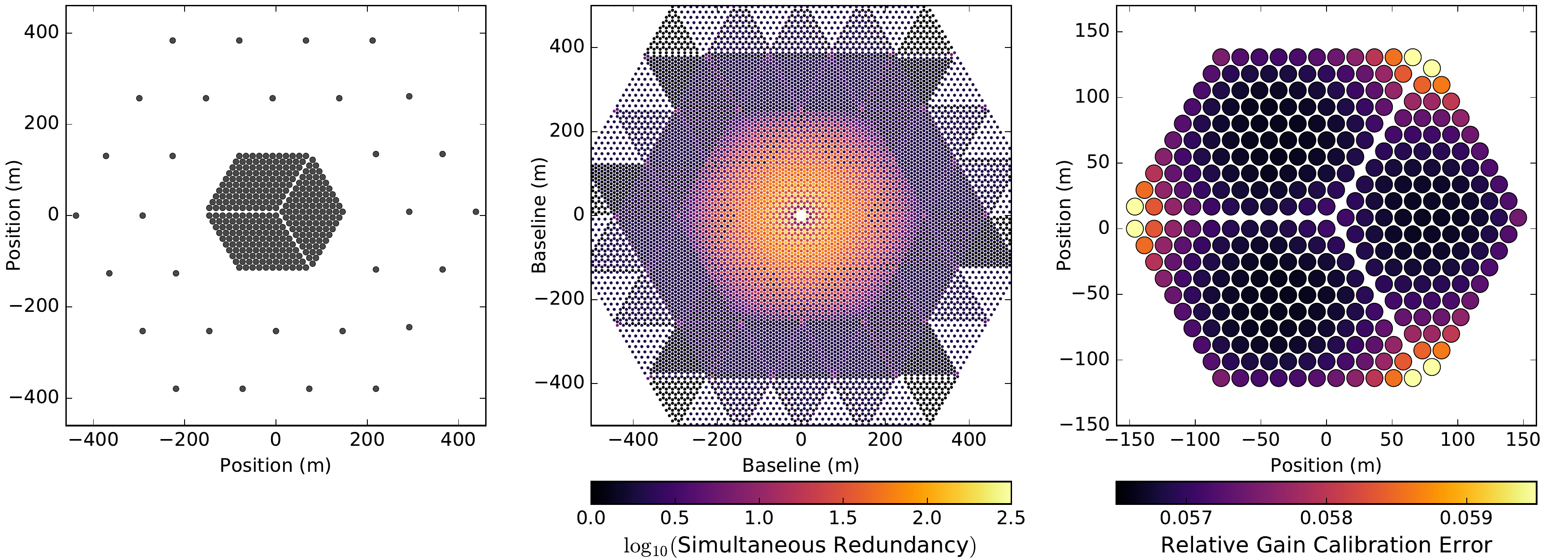}
	\caption{HERA's elements are divided between a 320-element, hexagonally-packed core and 30 outriggers (left). This produces instantaneous
$uv$ coverage at triple the element packing out to 250$\lambda$ at 150\,MHz, supressing grating lobes in the synthesized beam (middle). All 350 elements can be redundantly calibrated 
using the \citet{liu_et_al2010} technique, yielding calibration errors that are a small fraction of the residual noise per antenna (right).  See \cite{dillon_parsons2016} for discussion.}
		\label{fig:arrayConfig}
\end{figure*}

HERA's 320 core elements are arranged in a compact hexagonal grid, split into three displaced segments to cover the $uv$-plane with sub-aperture sampling density (see Fig.~\ref{fig:arrayConfig}). Splitting the core into three sections triples the density of instantaneous sampling of the $uv$-plane, improving HERA's mapmaking ability.  Furthermore, the core is supplemented by 30 additional outrigger elements to tile the $uv$-plane with instantaneously complete sub-aperture sampling out to 250$\lambda$ and complete aperture-scale sampling out to 350$\lambda$ (at 150\,MHz). 
This sub-aperture sampling strategy suppresses grating lobes in the instrument's point spread function and provides information for calibrating and correcting direction-dependent antenna responses \citep{dillon_parsons2016}.
Even with the sub-aperture dithering and long baselines, the design is sufficiently redundant to take advantage of redundant baseline calibration technique and robust to the failure of individual elements.
Resulting calibration errors range from $\sim 5\%$ (in the core) to $\sim 10\%$ (for outriggers) of the residual fractional noise per antenna after averaging. 
The split core increases the gain errors on the core elements by only $\sim 0.1\%$ from a solid hexagonal configuration. HERA's compact and redundant design allows it to fully utilize PAPER's robust approach to calibration and foreground mitigation, but it also provides for excellent imaging capability that can be leveraged to suppress foregrounds and improve access to the 21\,cm reionization signal.

\subsection{Analog Signal Path}
\label{sec:analog}
HERA's analog signal path from 50--250 MHz with the full performance EOR band of 100--200 MHz, as shown in Figure~\ref{fig:cosmos} and discussed in \S\ref{sec:freqs}, emphasizes spectral smoothness and robustness.  Although the low-frequency sky has an intrinsically bright signal level, in the desired ``cold patches'' the sky temperature is only about 30 K at the upper frequency end, so relatively low receiver temperatures are desired.  Note that at the low end of the band (50 MHz), the sky temperature is greater than 1000 K and across the EOR band it ranges from about 100--400 K.
With receiver temperatures of 30--100 K readily achievable with ambient temperature electronics however
HERA's low-noise amplifiers (LNAs) may be inexpensive, passively cooled components integrated into the
antenna feed.  

As part of the delay specification,
HERA's signal path emphasizes careful impedance matching between each component to minimize signal reflections that can worsen foreground leakage for the delays of interest.  This means that the emphasis is to reduce the delay-value of a reflection, which allows for a design degree of freedom.  One such design element is to house the analog-to-digital converters (ADCs) in ``nodes'' near the antenna elements
to limit the total analog path length.  This length is set to be 35~m, such that the round-trip delay corresponds to about 0.15~h~Mpc$^{-1}$, which remains in the wavenumbers  near the wedge.  The project is also examining the impact of having a range of varying analog path lengths.

The Phase 2/3 analog signal path comprises the front-end module (FEM) at the feed, the post-amplifier module (PAM) at field-deployed nodes, and the $\sim 35$ m coaxial cable in between.  Also included are the monitor and control aspects to control and measure its state.

\noindent
{\bf Front-End Module (FEM): }
The Front-End Module consists of an LNA as the very first stage in the chain for each polarization.  This component for the existing feed design will be a differential low noise amplifier which consists of two sets of LNAs feeding each dipole arm followed by a balun (or hybrid coupler).  The output from this point is then all single ended and mostly matched to 50~$\Omega$ impedance, excepted as needed to drive the Phase 1 75~$\Omega$ cables.  While it is desirable to be sky noise dominated in the HERA band, meaning for the first stage, the best achievable noise match must be obtained, it must not come at a too great cost to the power match since reducing the ripple across the band is absolutely essential.  This trade-off will be observed carefully to ensure the FEM is both low-noise and well-matched.  This stage is followed by carefully matched filters and further amplifications of the signal.  In order to ensure very low reflection on the line a large matching attenuator will be used at each end of the cable after the bias-Tee circuit. 

The FEM will also include a number of useful signal conditioning operations.  One of these will be to have a phase switch for each polarization as early as possible in the signal chain.  The switch will offer very low phase imbalance and will be controlled by differential lines.  The Walsh functions (orthogonal phase switching signals) will be generated in the SNAP board (Sec. \ref{sec:digital}) and used to control FEMs in the field.  

Another useful feature of the FEM will be an integrated ``Dicke'' switching radiometer similar to the EDGES experiment \citep{2012RaSc...47.0K06R}.  The circuit provides the capability of switching between the output of the antenna, a calibrated noise source and a 50~$\Omega$ ambient load.  The antenna measures only a fraction of the sky brightness temperature, determined by the matching between the antenna and receiver.  Three spectrometer measurements (antenna, ambient load and noise source) as well as additional lab measurements of its microwave performance and noise-parameters of the FEM prior to deployment then allow good calibration of the output spectrum. An on-board temperature sensor near the noise source will also aid with the calibration of the receiver chain.  The FEM will be housed in a rugged 5$\times$5$\times$10 cm unit which is water and dust resistant.

\noindent
{\bf Post-Amplifier Module (PAM): }
The Post-Amplifier Modules consists of further amplification and filtering of the RF signals received from the FEMs.  They are designed to provide the anti-aliasing filtering (DC-250 MHz) prior to the ADCs on the SNAP board.  These modules also feed the DC voltage supply to the FEMs on the antennas and relay the control and monitoring signals using a 1-Wire interface.  Each PAM will provides a controllable digital attenuator to allow the input levels into the ADC to be set.  All the controllable circuits such as RF switches and attenuators will have a unique address.  It is envisaged that most of the lower level control signals which are not data critical will be achieved though the node computer.  The phase switch, however, is a data critical control signal sent directly by the SNAP boards.  Each PAM will be rack mountable (3U, 5HP) and will control a single antenna containing 2 polarizations. 

\subsection{Digital Signal Path}
\label{sec:digital}
\begin{figure*}
\centering
\includegraphics[width=0.9\textwidth]{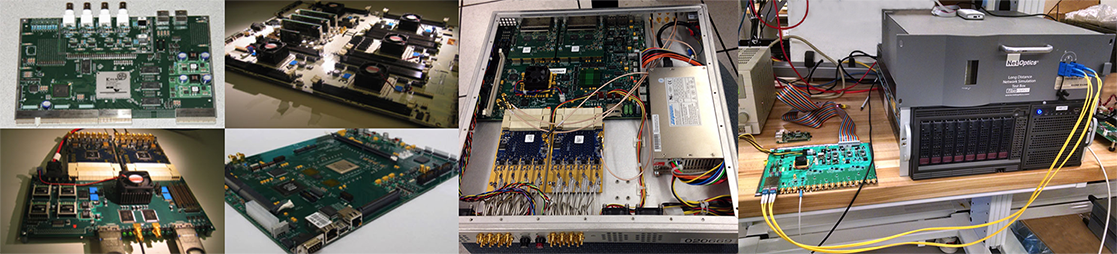}
\caption{
Six generations of CASPER digital signal processing (left to right) culminating in the SNAP board (right, along with the long-haul fiber-link test setup).
By preserving its design tools, signal processing libraries, and interface code between hardware generations, CASPER hardware and a modular architecture have enabled the PAPER correlator to be easily upgraded for HERA \Mycitep{parsons_et_al2006,parsons_et_al2008}.
}\label{fig:hardware}
\end{figure*}

Although in the past, correlator development has been one of the most expensive and complex aspects of building a large array or radio interferometers, this is no longer generally the case.
The Collaboration for Astronomy Signal Processing and Electronics Research (CASPER; \citealt{parsons_et_al2006},  {casper.berkeley.edu}) is a community of astronomers and engineers who work to reduce the cost and complexity of radio astronomy signal processing systems through the development of open-source, general-purpose hardware and software.
CASPER currently has several hundred members at 73 institutions, and has developed six generations of FPGA-based signal processing hardware, shown in Figure~\ref{fig:hardware}.
PAPER has applied CASPER technology to develop and deploy new correlators annually for five years running, each quadrupling the computational capacity of its predecessor.
Key to the upgradability of the PAPER correlator is the use of modular processing engines, and industry-standard digital interconnect based on off-the-shelf Ethernet switches \Mycitep{parsons_et_al2008} to perform the antenna/frequency data transpose required by FX correlators.

HERA will maintain both PAPER's well-proven digital system architecture along with the simple scheme of real-sampling and channelizing the entire analog passband at once.
However, in order to meet the 35\,m specification for maximum analog signal path length, as well as ensure future scalability of the system, HERA-350 will adopt an architecture of field-deployed amplification, digitization, and channelization nodes, building on MWA and Allen Telescope Array \citep{2009IEEEP..97.1438W} heritage.
Digital data streams from multiple nodes will converge to a container adjacent to the HERA array, from which they will be routed to a central processor building where correlation and further processing will take place.
This section will step through the digital signal path, shown in Figure~\ref{fig:overallBlockDiagram}, organized by location:  node, container, and processing building.

\vspace{.1in}
\noindent{\bf Node}. 
HERA-350 employs RFI-tight node enclosures that each contain the final gain (PAM) and digitization stages for signals from 12 antennas, along with power supplies, cooling, sensors, and a small server for monitor/control. 
This has led to the development of a new CASPER platform based around a Xilinx Kintex 7 FPGA, which incorporates on-board ADCs: the Smart Networked Analog-Digital Processor (SNAP\footnote{ {https://casper.berkeley.edu/wiki/SNAP}}
, shown in Fig. \ref{fig:hardware}).
SNAP is an inexpensive and flexible ``Analog in, 10 GbE out'' device which uses a pair of industry-standard SFP+ modules for output over either copper or optical fiber cables.
Co-designed by UC Berkeley and NRAO, the SNAP board will serve as both the digitizer and F-engine in HERA's FX correlator architecture.
Each SNAP board digitizes and channelizes a 0--250 MHz band for 6 input signals (3 antennas, dual-polarization), with a complete node containing 4 SNAP boards.

Following digitization and channelization, a $\sim$200-MHz band of runtime-selectable channels is output as a UDP stream over optical fiber.
Coarse Wavelength Division Multiplexing (CWDM) technology allows the 10 Gb/s Ethernet streams from the four SNAP boards in a node to share a single fiber, which is routed to a central container, and on to the correlator.
In this configuration, the fiber from each node is input directly into an Ethernet switch using a single QSFP+ 40GBASE-LR4 transceiver, which provides built-in multiplexing and demultiplexing capabilities and may be operated as 4 independent transceivers.
Laboratory tests have successfully demonstrated robust transmission from the SNAP board through CWDM multiplexers and 10 km single mode fiber optic cable, into a 10 GbE switch using inexpensive, commercially available optical transceivers (see far right of Fig. \ref{fig:hardware}).

While some applications may utilise SNAPs in multiple-board chassis, for HERA SNAP is housed in a field-deployable RFI-tight box.
The node enclosure itself is an RFI-tight housing that can handle the signal paths for up to 12 antennas.
The enclosures are placed throughout the array to minimize the total number constrained by the 35\,m maximum cable length. 
For outrigger antennas the delay specification is not critical, so the cable-length to those may be as long as practical.
In total, including outriggers, the HERA-350 design demands 34 nodes, or an average of just over 10 antennas/node.

The node uses forced-air, earth-coupled cooling similar to that implemented for the Allen Telescope Array \citep{2009IEEEP..97.1438W}.
An AC-synchronous motor continually blows air through 30\,m of underground 150\,mm PVC pipe, which vents through the enclosure from bottom to top.
The underground pipe is at least 1\,m deep, so that the emerging air is very thermally stable on seasonal time constants.
The PAM is the lowest component since it is the most sensitive and also generates the least heat.
SNAP boards, monitor and control hardware, and power supplies will be positioned above the PAMs.
Remote monitoring capabilities will include node temperature, air-flow and power, with remote control of power to the node's various active components.

\vspace{.2in}
\noindent{\bf Support Container}.
HERA's support container houses two significant subsystems adjacent to the array.
The first is a timing subsystem that maintains a GPS-disciplined oscillator and distributes timing signals to the nodes.
These signals comprise a sampling clock (or other frequency reference from which the sampling clock may be derived) and 1 pulse-per-second (PPS) synchronization pulses.
The timing subsystem will distribute clock and synchronization signals over fiber and, depending on cost and performance, will utilize either industry-standard distribution solutions, or the White Rabbit protocol \Mycitep{Moreira_WR}.

The second subsystem is a passive fiber optic patch panel that couples the optical network from the nodes into the 192-filament optical fiber bundle that is routed to the Karoo Array Processing Building.

\vspace{.2in}
\noindent{\bf HERA and the KAPB}.
Approximately 10\,km (by fiber path) from the array support container is the Karoo Array Processor Building (KAPB), the purpose-built facility to house radio astronomy computing resources in the remote desert introduced in Section~\ref{sec:site}.
A fiber optic bundle from the HERA container enters the KAPB and is patched into local fiber cables, which terminate in QSFP+ optical transceivers.
These QSFP+ transceivers are input to a 64-port QSFP+ 10/40 GbEthernet switch -- such devices are readily available on the commercial market today, at relatively low cost.
The Ethernet switch forms the core of the HERA correlator's data interconnect, and provides the antenna-frequency data transpose which allows cross-multiplication of antenna signals to be easily parallelized by frequency over many compute nodes performing cross-multiplication (``X-Engines'').

These compute nodes are anticipated to be 30 x86 servers, each hosting a pair of Nvidia Graphics Processing Units (GPUs) and two dual 10GbE network interface cards.
This estimate assumes an identical configuration to the current PAPER X-Engine hardware, and conservatively assumes that the computational capacity of GPU accelerator cards will double once prior to purchasing these servers.
No improvement in data transfer speed from CPU to host is assumed.

Other X-Engine realizations are under active evaluation, including a solution involving AMD GPUs and the highly-optimized cross-multiplication kernels developed for the CHIME Pathfinder array \Mycitep{Denman_Chime_X}, and FPGA-based implementations involving both CASPER, and commercially available, platforms.

Output data from the correlator are written to the data storage system described in the following section.

\subsection{Data Processing and Management}
\label{sec:data}
\begin{figure}
\centering
\includegraphics[width=0.48\textwidth]{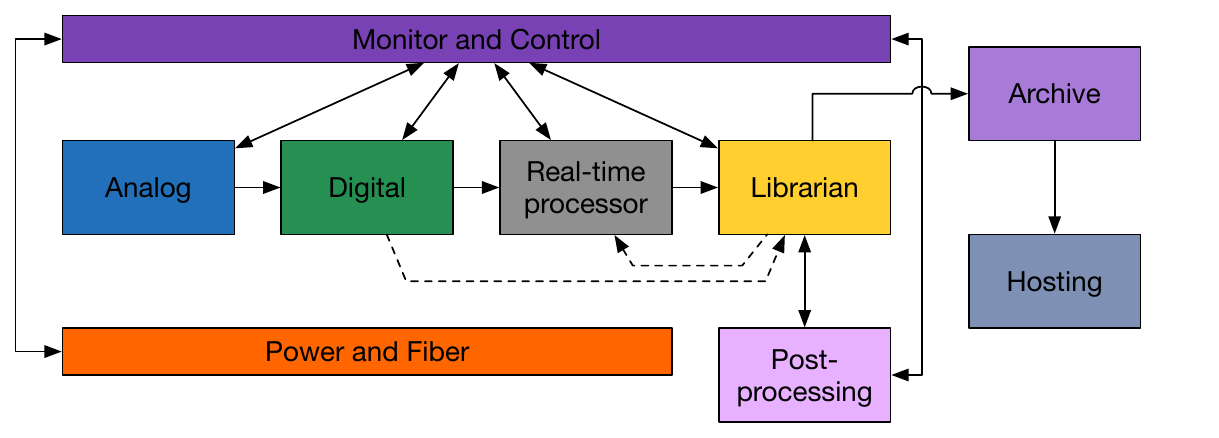}
\caption{
High-level architecture 
}\label{fig:sysover}
\end{figure}

\noindent The HERA correlator generates $\sim4$~TB of raw data per 12-hour observing day.  These data are transferred to an on-site cluster (the Real Time Processor, RTP) where they are calibrated on the basis of redundancy \citep{zheng_et_al2014} and averaged in time and frequency to the limit possible without information loss. Raw products are 
archived in an on-site 1.5 PB storage array; reduced size products are transferred via the internet to the NRAO Data Archive. The RTP supports multiple data volume reduction schemes including delay-delay-rate filtering \citep{parsons_etal2014}, sub-array selection, and baseline dependent averaging. 
The data cataloging system, known as the ``librarian'' catalogs files, coordinates data transfers, and cross-references data to observation meta-data.
The high-level architecture is shown in Figure~\ref{fig:sysover}

NRAO will host the archiving of HERA calibration products and images within the existing NRAO EVLA Next Generation Archive System (NGAS) at the Domenici Science Operations Center (DSOC). The NGAS software provides indexing via instrument defined keywords and supports public and private retrieval of both raw and processed products via web based searches with access controlled by the NRAO user database.

Routine data inspection and lightweight analysis tasks is performed on a processing cluster within the DSOC and includes access to a 1PB high speed Lustre filesystem.  For bulk application of processing-intensive data pipelines to HERA data NRAO manages the distribution of load between the in house cluster, NSF funded XSEDE resources ( {www.xsede.com}) and if necessary costed access to Amazon Web Services (AWS) cloud computing via AWS's spot market ( {aws.amazon.com/ec2/spot})

\subsection{Analysis Pipelines} 
\label{sec:software}

\noindent HERA builds on the rich legacy of PAPER and MWA software and database systems 
developed for field operations, data analysis, and simulation.  Examples range from
the strictly versioned and unit-tested packages for field-deployed systems (e.g. the correlator, real-time processing, and
monitor/control systems) to loose collections of scripts written for exploratory analysis.
Software packages that support HERA analysis are open source, publicly 
hosted\footnote{Including  {github.com/AaronParsons/aipy},  {github.com/JeffZhen/omnical}, 
 {github.com/MiguelFMorales/FHD},  {github.com/MiguelFMorales/eppsilon},  {github.com/jpober/21cmSense},
 {github.com/nithyanandan/PRISim}, and  {https://github.com/BradGreig/21CMMC}.}, revision controlled, and unit-tested.  
These include OMNICAL, a complete package for redundant baseline calibration;
Astronomical Interferometry in Python (AIPY), a set of
tools and file-format interfaces for reading visibilities, calibrating,
rephasing, imaging, and deconvolution; Fast Holographic Deconvolution
(FHD), a purpose-built tool for imaging, calibration, and foreground
forward-modeling and subtraction \citep{sullivan_et_al2012}; Precision Radio Interferometry Simulator 
(PRISim), a package for accurately simulating wide-field interferometric observations;
21cmFAST \citep{mesinger_et_al2011}, a fast, semi-numerical 21~cm signal simulator,
and 21cmSense \citep{pober_et_al2014}, a tool for forecasting power spectrum sensitivity.
Other project code is aggregated and revision
controlled in a public repository with separate sandboxes for each developer to ensure that
HERA members have up-to-date copies of all project code to facilitate sharing and debugging.
Such code includes the PAPER pipeline for foreground filtering and estimating power spectra from
visibility data,
the MWA power spectrum analysis codes---\eppsilon\ \citep{hazelton_et_al2016} and the 
empirical covariance technique of \citet{dillon_et_al2015}---as well as
machine-learning-based source finding, verification, and removal tools \citep{2016arXiv160703861C,2016ApJ...825..114J,beardsley_et_al2016}.

New software development is focusing on integrating and improving the MWA and PAPER power spectrum and foreground removal pipelines,
developing a monitor and control software interface and database for recording instrument metadata,
extending, with support from Scuola Normale Superiore, semi-analytic and numerical models 
of the 21~cm signal for robust parameter estimation, and developing
machine learning interpolations of simulations for joint Monte Carlo estimation of cosmological and astrophysical parameters.

\section{Schedule and Status}
\label{sec:status}

\noindent On-site construction is proceeding in stages, starting from the
initial 19 elements currently in place.    Elements 20--37
are currently under construction, to be completed in 2016.  
 These will be used in conjunction with the thoroughly characterized extant PAPER signal path and
processing hardware for a very low risk initiation of scientific observations.
As elements 38 to 128 are installed in 2017, they can
immediately be placed within the array and be used for observing.  

\begin{figure}
	\includegraphics[width=0.48\textwidth]{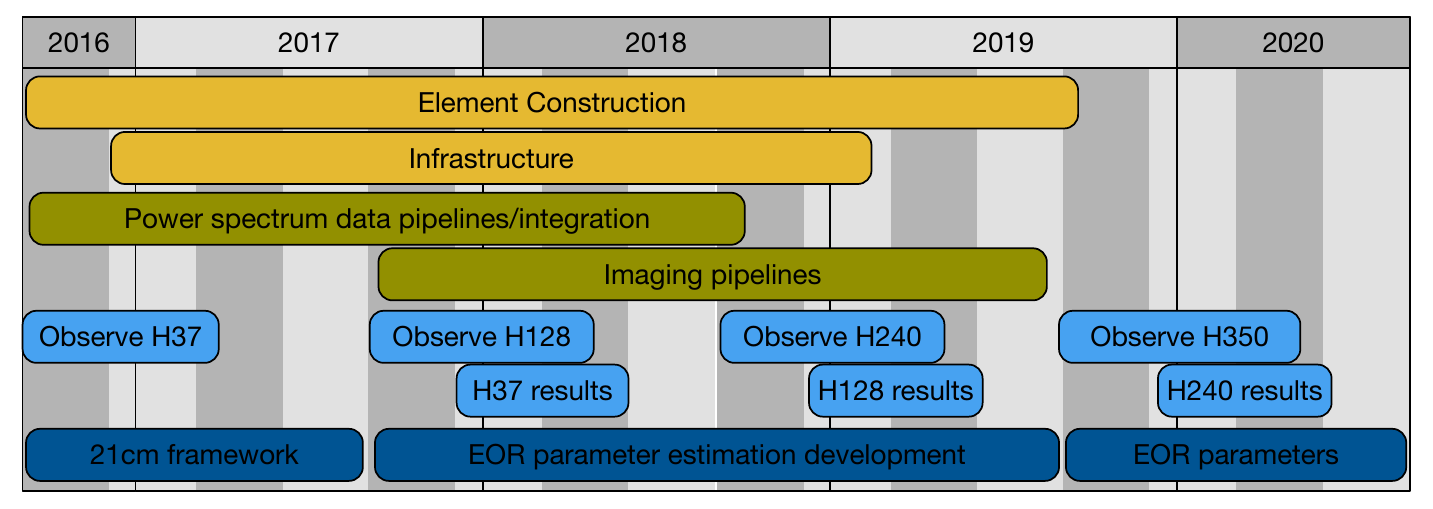}
	\caption{Timeline of HERA construction, analysis development, observation, and scientific output.  Activities after 2016 are contingent upon funding.}
	\label{fig:timeline}
\end{figure}

During this period, infrastructure for the new node-based system will be installed
and tested with the first elements beyond 128.  After the HERA-128 observing
season, the full array will be transitioned to use HERA's new hardware infrastructure.  
In this same time frame, the existing PAPER processing
container will be moved to the edge of the array to house the timing sub-system
and fiber optic splice cabinet while the correlator will move to the KAPB.
 The overall plan is shown in Figure~\ref{fig:timeline}.

The optimal EoR observing window is from September to April, with power spectrum limit results appearing about one year later.  Concurrent technique development and deployment will be on-going.
A breakdown of activities is shown below.
\begin{itemize}[leftmargin=0.7in]
\item[2016/17:] Observe with H37.  Real-time data pipeline, delay-space power spectrum (DSPS) pipeline, FHD pipeline.  21\,cm framework for incorporating with other probes. Construct H128.
\item[2017/18:] Observe with H128.  Real-time calibration pipeline.  DSPS/FHD/global sky model integration. Snapshot imaging pipeline.  EoR parameter estimation development.  H37 results.  Construct H240.  {\em Data products:  power spectrum, Stokes I maps.}
\item[2018/19:] Observe with H240.  Foreground-filtered imaging pipeline.  EoR parameter estimation development.   H128 results. Construct H350.  {\em Data products:  power spectrum, Stokes IQUV maps, foreground image cube.}
\item[2019/20:] Observe with H350.  EOR parameter estimates.  H240 results.  {\em Data products:  power spectrum, global sky model IQUV, snapshots, foreground-filtered image cube.}
\item[2020+:] Observe with H350.  H350 results.  {\em Data products:  power spectrum, global sky model IQUV, snapshots, foreground-filtered image cube.}
\end{itemize}

\section{Conclusion}
\label{sec:conclusion}
The first generation instruments are beginning to provide constraints on some models of the reionization of the universe, as well as developing the key algorithms and 
comprehension to enable detection of the power spectrum of the Epoch of Reionization.   
In the past three years, we have developed the EoR window paradigm for isolating foreground systematics, implemented novel
calibration and power spectrum analysis pipelines, made precision measurements of astrophysical foregrounds, and published deep power 
spectrum limits that constrain heating in the early universe.
However, as shown in Figures \ref{fig:Sensitivities} and \ref{fig:paramConstraints} and Table \ref{tab:signif}, using proven methods these instruments are not likely to make a robust detection or enable its characterization as a function of redshift or astrophysical parameters.  We have drawn from this development to design and develop a purpose-built array to detect (or, if current arrays succeed, provide a robust confirmation of) the signature of the power spectrum of the Epoch of Reionization.  Since it is designed for a specific experiment, HERA's optimization allows for a substantial increase in sensitivity to enable precise constraints of EOR astrophysics and a broad range of high-impact secondary science.

HERA will use the delay-spectrum approach to detect and characterize the EOR across its full redshift range at spatial scales that inform cosmology and astrophysics in the early universe.  The first 19 elements are on the ground and being used with extant components from PAPER, and construction has begun on another 18 to bring the array to 37 elements.  With 37 elements, HERA should have sufficient sensitivity to detect the peak of reionization at wavenumbers greater than 0.15 hMpc$^{-1}$ with a season's worth of observing.  Building out to the full 350 elements should enable HERA to fully characterize the EOR power spectrum and potentially begin to directly image this epoch over a portion of the sky.  These results are a necessary component to enable future arrays like the Square Kilometre Array (SKA; {\em e.g.} \citealt{2013ExA....36..235M,2015arXiv150903312G}) to do detailed mapping of structures over the entire sky.

\noindent
{\bf Acknowledgements:}  
This work was supported by the U.S. National Science Foundation (NSF) through awards AST-1440343 and AST-1410719.
ARP acknowledges support from NSF CAREER award 1352519.
AL acknowledges support for this work by NASA through Hubble Fellowship grant \#HST-HF2-51363.001-A awarded by the Space Telescope Science Institute, which is operated by the Association of Universities for Research in Astronomy, Inc., for NASA, under contract NAS5-26555.
This research was completed as part of the University of California Cosmic Dawn Initiative. AL, ARP, and SRF acknowledge support from the University of California Office of the President Multicampus Research Programs and Initiatives through award MR-15-328388

\bibliographystyle{apj}
\bibliography{biblio}

\end{document}